\documentclass[
  journal=pla,
  manuscript=article-type,
  year=,
  volume=,
]{cup-journal}

\usepackage{amsmath}
\usepackage{amsfonts}
\usepackage[nopatch]{microtype}
\usepackage{booktabs}
\usepackage{caption}
\usepackage{graphicx}
\usepackage{pdflscape}
\usepackage{array}
\usepackage{booktabs}
\usepackage{multirow}
\usepackage{tabularx}
\usepackage{subcaption}
\usepackage{pifont}

\title{Euclidean Ideal Point Estimation From Roll-Call Data via Distance-Based Bipartite Network Models}

\author{Seungju Lee}
\affiliation{Department of Statistics and Data Science, Yonsei University. Republic of Korea.}

\author{In Kyun Kim}
\affiliation{Department of Statistics and Data Science, Yonsei University. Republic of Korea.}
\alsoaffiliation{Department of Political Science and International Relations, Seoul National University. Republic of Korea.}

\author{Jong Hee Park}
\affiliation{Department of Political Science and International Relations, Seoul National University. Republic of Korea.}
\email[Jong Hee Park]{jongheepark@snu.ac.kr}

\author{Ick Hoon Jin}
\affiliation{Department of Statistics and Data Science, Yonsei University. Republic of Korea.}
\email[Ick Hoon Jin]{ijin@yonsei.ac.kr}


\begin{document}

\begin{abstract}
Conventional ideal point models rely on Gaussian or quadratic utility functions that violate the triangle inequality, producing non-metric distances that complicate geometric interpretation and undermine clustering and dispersion-based analyses. We introduce a distance-based alternative that adapts the Latent Space Item Response Model (LSIRM) to roll-call data, treating legislators and bills as nodes in a bipartite network jointly embedded in a Euclidean metric space. Through controlled simulations, Euclidean LSIRM consistently recovers latent coalition structure with superior cluster separation relative to existing methods. Applied to the 118th U.S. House, \textcolor{black}{the model provides competitive predictive performance while yielding bill embeddings that clarify cross-cutting issue alignments.} The results show that restoring metric structure to ideal point estimation provides a clearer and more coherent inference about party cohesion, factional divisions, and multidimensional legislative behavior.
\end{abstract}

\section{Introduction}\label{sec:intro}

Ideal point estimation from roll-call voting data has become a foundational tool in legislative studies, enabling systematic analysis of partisan polarization, ideological trajectories, and coalition dynamics \citep{Poole:2007, Clinton2004, mccarty2016}. The dominant methods—NOMINATE \citep{Poole:2007, voteview2025} and Bayesian Item Response Theory (BIRT)  \citep{Clinton2004, pscl2024}-successfully recover ordinal ideological relationships and achieve high vote classification accuracy. Recent work expands these frameworks by incorporating domain-specific \textcolor{black}{ideal-point models or multidimensional extensions} \citep{moser2021multiple, marble2022structure, binding2023non, lipman2025explaining}, integrating text-based information \citep{kim2018estimating}, or accommodating both monotonic and non-monotonic response functions \citep{lei2025novel}. 

Yet despite these advances, existing models share a structural limitation: the utility functions underlying NOMINATE and BIRT do not induce proper metric distances. Their Gaussian and quadratic specifications violate the triangle inequality, the defining property of metric spaces. Although this issue is mathematically straightforward, its consequences for political analysis are substantive. When distances lack metric validity, dispersion statistics used to assess party cohesion become non-comparable, and clustering analyses may generate configurations in which A is ``close” to B, B to C, yet A remains far from C. As contemporary legislatures exhibit growing intra-party heterogeneity \citep{Clarke:2020, Harbridge:2023}, researchers increasingly rely on spatial distances to characterize factional structure, coalition tendencies, or ideological spread—applications for which \textcolor{black}{utility functions that do not induce metric distances (“non-metric utilities”)} provide an unstable foundation. 

\textcolor{black}{Recent work in the latent space literature highlights that different latent-space geometries can organize relational data in distinct ways. In particular, inner-product formulations capture directional similarity between latent vectors, whereas Euclidean distance models represent affinity through spatial proximity. These alternative geometries can lead to different representations of coalition structure even when the resulting models achieve similar predictive accuracy.}

\begin{figure}[h]
    \centering
    \begin{minipage}{0.45\linewidth}
        \centering
        \includegraphics[width=\linewidth]{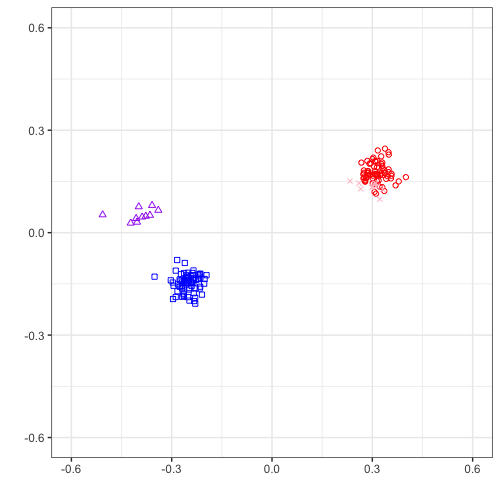}
        \text{(a) Bayesian IRT}
    \end{minipage}
    \begin{minipage}{0.45\linewidth}
        \centering
        \includegraphics[width=\linewidth]{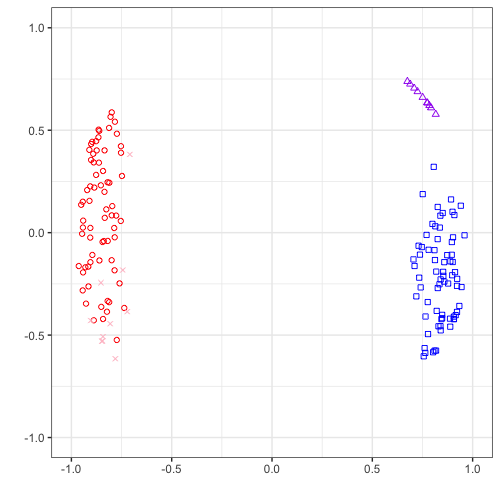}
        \text{(b) NOMINATE}  
    \end{minipage}
    \caption{Coalition-cluster distortion in non-metric ideal point methods. Ground truth: \textbf{four distinct coalitions} (marked by colors and shapes)—Party A majority (70, blue squares) and minority (10, purple triangles); Party B majority (70, red circles) and minority (10, pink crosses). The 400-bill agenda includes 200 partisan bills (party-line division) and 200 faction-specific bills (50 per faction). Despite this clear structure, both BIRT (a) and NOMINATE (b) compress the four factions into approximately two party blocs, with faction markers overlapping substantially within parties.}
    \label{fig:coalition_cluster_distortion}
\end{figure}
\textcolor{black}{These geometric differences have practical consequences. Figure \ref{fig:coalition_cluster_distortion} illustrates this issue using simulated data with four distinct coalitions.} Both NOMINATE and BIRT systematically compress factions into two party blobs. \textcolor{black}{While both methods achieve high vote classification accuracy, the resulting spatial representations do not clearly recover the four underlying factions.}

Recent work has recognized these limitations. \citet{Shin2024} proposed an $L_1$-norm model using Manhattan distance: $U(x,y) = -\sum_k w_k|x_k - y_k|$, which satisfies the triangle inequality and provides a valid metric \textcolor{black}{where $x$ and $y$ denote the latent ideological positions of a legislator and a policy alternative respectively in the spatial model.} However, Manhattan distance creates interpretive difficulties for spatial visualization: the distance between two points depends on their axis-aligned path rather than the direct geometric separation, complicating intuitive understanding of ideological proximity. 

In this paper, we propose using Euclidean distance ($L_2$ norm) $d(x,y) = \sqrt{\sum_k (x_k - y_k)^2}$, which provides both valid metric properties and intuitive geometric interpretability of spatial models. Euclidean distance satisfies the triangle inequality through the Cauchy--Schwarz inequality, provides the familiar notion of ``straight-line'' distance between points, and admits natural geometric interpretation of clusters as spatially coherent regions.

Our approach builds on recent network-based measures of political ideology. \citet{Barbera2015} develops a Bayesian spatial following model for Twitter data\textcolor{black}{, but} its quadratic distance formulation ($-\lVert x_i - x_j\rVert^2$) inherits the same non-metric limitation. Similar work by \citet{lo2025statistical} treats cosponsorship as a bipartite network and estimates latent positions via a mixed-membership blockmodel, \textcolor{black}{where legislators and bills are represented as membership vectors on a simplex. These mixed-membership embeddings capture probabilistic affiliation with latent groups but do not retain the Euclidean distance structure assumed by spatial voting models.} In contrast, we adopt the Latent Space Item Response Model \citep[LSIRM]{jeon:2020}, which treats roll-call data explicitly as a bipartite legislator–bill network and embeds both node types jointly in a Euclidean metric space. This symmetric formulation enables bills to serve as interpretive anchors \textcolor{black}{for the latent dimensions while maintaining metric consistency for clustering and coalition analysis.}

In the following, we provide theoretical and empirical evidence on how different latent-space geometries affect the interpretation of legislative voting behavior. We show that common ideal point utility specifications do not define metric distances, with implications specifically for distance-based interpretations. We then demonstrate that joint legislator--bill embedding aids substantive interpretation through bill locations, \textcolor{black}{and that Euclidean LSIRM provides predictive performance competitive with BIRT while clarifying cross-cutting issue alignments.}
 
\textcolor{black}{Section 2 discusses the geometric properties of common ideal point utility functions and clarifies the role of metric distance in spatial interpretation.} Section 3 introduces the bipartite network framework and Euclidean LSIRM specification. \textcolor{black}{Section 4 presents simulations comparing Euclidean and inner-product formulations in recovering coalition structure.} Section 5 applies the method to the 118th House, showing how bill embeddings serve as interpretive anchors for coalition structure. Section 6 concludes with implications and extensions.

\section{Metric Properties of Utility Functions}\label{sec:theory}
\subsection{Proximity Voting Theory and Metric Requirements}

The theoretical foundation for spatial voting models rests on the proximity voting framework developed by \citet{downs1957economic} and formalized by \citet{davis1970expository}. Under this framework, each voter $i$ possesses an ideal point $x_i$ in a policy space, and evaluates alternatives \textcolor{black}{($y$)} by their distance from this ideal point. For a policy located at position $y$, the voter's utility follows:
\begin{equation}
U_i(y) = -d(x_i, y) + \epsilon_i,
\label{eq:proximity_utility}
\end{equation}
where $d(\cdot, \cdot)$ represents a distance function and $\epsilon_i$ is a random utility component. The voter prefers alternative $A$ over alternative $B$ when $U_i(A) > U_i(B)$; absent stochastic shocks, this occurs when $d(x_i,A)<d(x_i,B)$, so $A$ is closer to the voter's ideal point than $B$. With shocks, the comparison also depends on $\epsilon_{iA}-\epsilon_{iB}$: under $U_i(j)=-d(x_i,j)+\epsilon_{ij}$, $U_i(A)>U_i(B)$ holds when $d(x_i,A)-d(x_i,B)<\epsilon_{iA}-\epsilon_{iB}$.

The proximity voting framework implicitly assumes that $d(\cdot, \cdot)$ constitutes a proper distance metric. Three properties are essential for the theoretical coherence of this framework. First, symmetry requires $d(x,y) = d(y,x)$: the ideological distance from position $x$ to position $y$ equals the distance from $y$ to $x$. Without symmetry, the model would imply that moving from a liberal to conservative position involves a different ideological journey than moving from conservative to liberal, contradicting the reciprocal nature of ideological proximity. Second, the triangle inequality requires $d(x,y) + d(y,z) \geq d(x,z)$ for all points: the direct distance between two positions cannot exceed the sum of distances through an intermediate position. Without this property, indirect ideological paths could be shorter than direct paths, violating basic intuitions about ideological space and enabling paradoxical configurations where three positions cannot be consistently arranged along ideological dimensions. Third, positive definiteness requires $d(x,y) > 0$ for $x \neq y$ and $d(x,x) = 0$: distinct positions must be separated by positive distance, while a position's distance from itself is zero. Without this property, distinct ideological positions could be indistinguishable, rendering the spatial representation meaningless.

These metric properties are not merely mathematical conveniences—they are fundamental to the substantive interpretation of proximity voting. Consider the triangle inequality: if legislator $A$ has ideal point close to position $x$, and position $x$ is close to position $y$, the triangle inequality guarantees that legislator $A$ cannot be arbitrarily distant from position $y$. This transitive consistency underlies claims about coalition formation: if moderate Democrats are close to centrist positions, and centrist positions are close to moderate Republicans, then moderate Democrats and moderate Republicans must be relatively proximate, enabling bipartisan coalitions. Without the triangle inequality, such transitive reasoning fails—proximity becomes intransitive, and the spatial metaphor loses coherent meaning.

\textcolor{black}{
Given these metric requirements, we now examine the utility formulations commonly used in spatial voting models, focusing on whether quadratic and Gaussian utility preserve the triangle inequality required for a metric notion of ideological proximity. This discussion concerns the theoretical utility formulation; issues related to how ideal point models are represented and estimated in practice are addressed separately separately in Section~\ref{sec:model}.
}

\subsection{The Triangle Inequality and Its Violation in Conventional Methods}

A distance function $d: \mathcal{X} \times \mathcal{X} \rightarrow \mathbb{R}_+$ defines a metric space if it satisfies four properties: (1) non-negativity: $d(x,y) \geq 0$ with equality if and only if $x=y$; (2) symmetry: $d(x,y) = d(y,x)$; (3) triangle inequality: $d(x,y) + d(y,z) \geq d(x,z)$ for all $x,y,z \in \mathcal{X}$; and (4) identity of indiscernibles. \textcolor{black}{The triangle inequality is particularly important: it ensures that distances are additive, comparable, and transitively consistent. Without this property, distance comparisons become difficult to interpret---for example, it becomes unclear whether one pair is ``twice as distant'' as another or whether distance-based clustering methods that assume metric structure are appropriate.}

\textcolor{black}{In terms of their theoretical specification, conventional spatial voting models typically employ utility formulations based on squared distance.} In BIRT \citep{Clinton2004}, legislators maximize quadratic utility derived from proximity voting: for each roll-call $j$ presenting a binary choice between ``Yea'' position $\zeta_j$ and ``Nay'' position $\psi_j$, legislator $i$ with ideal point $x_i$ votes ``Yea'' when $$-\lVert x_i - \zeta_j \rVert^2 + \eta_{ij} > -\lVert x_i - \psi_j \rVert^2 + \nu_{ij}.$$ \textcolor{black}{This formulation implies that utility decreases with the squared Euclidean distance between the legislator and policy positions. However, squared distance itself does not satisfy the triangle inequality and therefore does not define a metric distance.} Consider $d_Q(x,y) = (x-y)^2$ in one dimension. The triangle inequality requires $(x-y)^2 + (y-z)^2 \geq (x-z)^2$. Expanding $(x-z)^2 = [(x-y) + (y-z)]^2 = (x-y)^2 + 2(x-y)(y-z) + (y-z)^2$, the inequality reduces to \textcolor{black}{$2(x-y)(y-z) \leq 0$. This condition holds when $(x-y)$ and $(y-z)$ have opposite signs (or when one of them is zero), but fails when they have the same sign.}

In NOMINATE \citep{Poole:2007}, \textcolor{black}{spatial voting is often motivated 
using Gaussian utility:} 
$$U_i(\zeta_j) = \beta \exp\{-\frac{1}{2}\sum_k w_k^2(x_{ik} - \zeta_{jk})^2\}.$$ 
\textcolor{black}{The exponential transformation of squared distances alters the geometry 
implied by spatial proximity. To illustrate, consider the one-dimensional quantity 
$d_G(x,y) = -\exp\{-\tfrac{1}{2}(x-y)^2\}$. For three points $x=3$, $y=3$, and $z=-1$, 
we obtain 
$$d_G(x,y) = -\exp\{0\} = -1, \quad d_G(y,z) = -\exp\{-8\}, \quad d_G(x,z) = -\exp\{-8\}.$$
The triangle inequality would require $d_G(x,y) + d_G(y,z) \geq d_G(x,z)$, but 
$$-1 - \exp\{-8\} < -\exp\{-8\},$$ so the inequality is violated. Thus the geometry 
implied by Gaussian utility does not correspond to a metric distance.}

\textcolor{black}{Such configurations can arise when legislators are presented with 
policy alternatives that move policy in the same ideological direction but by different 
magnitudes. For example, suppose a legislator's ideal point lies near the status quo, 
and two amendments propose more moderate and more extreme versions of a policy change 
in the same direction. In this situation, the legislator, the moderate amendment, and 
the more extreme amendment lie along a single ideological line but at increasing 
distances from one another. Under squared-distance comparisons, this configuration 
can violate the triangle inequality even in a one-dimensional policy space.}

\textcolor{black}{While such violations can arise even in one-dimensional settings, 
their practical importance depends on the structure of the voting environment. In 
highly structured left--right settings, where alternatives are consistently ordered 
along a single ideological dimension, conventional models can recover the dominant 
partisan structure reasonably well. However, violations become more substantively 
relevant when the voting environment departs from this simple ordering---for instance, 
when both alternatives lie on the same ideological side (e.g., two competing 
left-leaning amendments), when procedural votes place both positions far from most 
ideal points, or when a unidimensional model is fit to an inherently multi-dimensional 
space.}

\textcolor{black}{When such configurations occur, the resulting non-metric geometry can 
complicate geometric interpretations of legislators' positions in the ideological space. 
For example, researchers often summarize party cohesion using the dispersion of 
legislators' positions within a party. If the underlying distance representation is 
non-metric, differences in dispersion may partly reflect how the utility formulation 
scales distances across the ideological space rather than only differences in voting 
cohesion. Similar considerations arise when clustering legislator positions to identify 
factions or when comparing ideological separations across policy dimensions. The concern we raise is structural rather than empirical: 
a non-metric distance function lacks the geometric guarantees---transitivity of 
separations, additivity through intermediate points, and cross-dimensional 
comparability---that underlie distance-based inference, regardless of how often 
violations are realized in a given dataset.}

\subsection{Euclidean Distance as a Metric}

Euclidean distance $d_E(x,y) = \sqrt{\sum_{k=1}^K (x_k - y_k)^2}$ satisfies the triangle inequality and defines a proper metric space. The proof follows directly from the Cauchy-Schwarz inequality: for any three points $x, y, z \in \mathbb{R}^K$, squaring $d_E(x,y) + d_E(y,z) \geq d_E(x,z)$ and expanding yields $d_E(x,y)d_E(y,z) \geq \sum_k (x_k-y_k)(y_k-z_k)$, which holds because $|\sum_k (x_k-y_k)(y_k-z_k)| \leq \sqrt{\sum_k (x_k-y_k)^2}\sqrt{\sum_k (y_k-z_k)^2} = d_E(x,y)d_E(y,z)$ by Cauchy-Schwarz.

This metric property has direct implications for legislative analysis. First, distances become cardinally interpretable: if $d_E(z_A, z_B) = 2 d_E(z_C, z_D)$, legislators $A$ and $B$ are genuinely twice as ideologically separated as $C$ and $D$ in their voting behavior. Second, distances become comparable across dimensions: we can meaningfully ask whether the first or second dimension generates greater polarization by comparing distance statistics, \textcolor{black}{because the metric structure allows these comparisons to be interpreted directly in terms of spatial separation in the latent space.} Third, clustering methods that assume metric distances—$k$-means, hierarchical clustering, silhouette analysis—\textcolor{black}{can be applied in a geometrically coherent way because pairwise distances satisfy the triangle inequality. For example, the silhouette coefficient measures cluster quality by comparing mean within-cluster distances to mean nearest-cluster distances, a computation that requires pairwise distances to be metrically consistent.} Fourth, the geometric interpretation aligns with mathematical properties: points that appear close in two-dimensional visualizations genuinely vote similarly, and spatial neighborhoods correspond to actual voting coalitions rather than distortions induced by non-metric utility functions.


\textcolor{black}{At the same time, metric embeddings are not free from scaling artifacts. As in other low-dimensional embedding methods such as multidimensional scaling or spectral embeddings, geometric patterns such as curvature or horseshoe structures may arise when complex preference structures are approximated in a limited number of dimensions \citep[e.g.,][]{Diaconis2008}. These limitations reflect the challenge of representing high-dimensional preference structures in a small number of spatial dimensions rather than a failure of the metric formulation itself.}
\section{Euclidean Distance Model for Bipartite Legislative Networks}\label{sec:model}

\subsection{Bipartite Network Framework}

Roll-call voting data constitute a bipartite network linking legislators to bills through binary votes (excluding abstentions). \textcolor{black}{Formally, let $G=(L,B,E)$ denote a bipartite graph where $L=\{1,\ldots,N\}$ represents legislators and $B=\{1,\ldots,M\}$ represents bills. An edge between legislator $i$ and bill $j$ is represented by the adjacency variable $y_{ij}$, where}
\[
y_{ij} =
\begin{cases}
1 & \text{if legislator } i \text{ votes ``yea'' on bill } j, \\
0 & \text{otherwise}.
\end{cases}
\]
\textcolor{black}{The resulting $N \times M$ adjacency matrix $Y=(y_{ij})$ corresponds directly to the roll-call voting matrix commonly analyzed in ideal point estimation. This representation treats roll-call voting as an affiliation network linking legislators to bills. Alternative network constructions are possible—for example, including status quo nodes or using a doubled network that treats ``yea'' and ``nay'' as separate edge types. We adopt the legislator--bill affiliation network because it corresponds directly to the observed roll-call matrix and enables joint embedding of legislators and bills in a shared latent policy space. In our formulation, nay votes are already incorporated through the complement of the modeled probability, so a doubled network would increase complexity without adding information.} 

\textcolor{black}{Latent variable models provide a natural framework for representing relational data of this type. In the latent space approach to network modeling \citep{Hoff:2002, Hoff:2007}, nodes are assigned positions in a low-dimensional space, and the probability of an edge depends on the relationship between their latent positions. Adapting this idea to the legislator--bill setting, let $z_i \in \mathbb{R}^K$ denote the latent position of legislator $i$ and $w_j \in \mathbb{R}^K$ the latent position of bill $j$.} The probability of a ``Yea'' vote depends on their spatial relationship:
\begin{equation}
g\left(\Pr(y_{ij}=1 \mid z_i, w_j)\right) = \mu + \alpha(z_i, w_j),
\label{eq:general_network}
\end{equation}
where $g(\cdot)$ denotes a link function and $\alpha(z_i, w_j)$ specifies how latent positions determine voting propensity. Three canonical specifications exist. The latent class model \citep[LCM, ][]{Holland1983, Nowicki:2001, Airoldi:2008, miller:2009, Morup:2011, Palla:2012, zhou2015infinite} and stochastic blockmodel \citep[SBM, ][]{Holland1983, Nowicki:2001, Airoldi:2008, daudin:2008, Karrer:2011, Vu:2013} sets $\alpha(z_i, w_j) = m_{u_i, v_j}$ for discrete group memberships $u_i, v_j \in \{1,\ldots,K\}$ and interaction matrix $M = (m_{k\ell})$, capturing stochastic equivalence where legislators in the same class vote identically. The latent eigenmodel \citep{Hoff:2007, young:2007, Hoff:2009, Li:2011, Hoff:2021} defines $\alpha(z_i, w_j) = z_i^{\top}\Lambda w_j$ for diagonal matrix $\Lambda$, representing voting through weighted inner products that extend principal component analysis to binary data. The latent distance model \citep{Hoff:2002, Schweinberger:03p307, Kemp:2006, Handcock:2007, Hoff:2007, latentnet2008, Raftery:2012, sewell2015latent, smith2019geometry} specifies $\alpha(z_i, w_j) = -\gamma \lVert z_i - w_j \rVert$, where voting probability decreases monotonically with spatial separation, operationalizing homophily through spatial proximity.

Conventional ideal point methods, however, treat this bipartite structure asymmetrically: legislators receive interpretable spatial positions (ideal points), while bills serve primarily as instruments for scaling, with bill parameters $(\alpha_j, \beta_j)$ estimated but not embedded in the same space. \textcolor{black}{In BIRT \citep{Clinton2004}, the probability of a ``Yea'' vote is typically written as}
\[
\text{logit}\,P(y_{ij}=1) = \alpha_j + \beta_j^{\top} x_i ,
\]
\textcolor{black}{where $x_i$ denotes the legislator's ideal point and $(\alpha_j,\beta_j)$ are bill parameters. Although this formulation is motivated by spatial proximity voting, the statistical representation of the model reduces to an interaction between legislator positions and bill loadings through the inner product $\beta_j^{\top} x_i$. As a result, bills are not represented as identifiable spatial locations comparable to legislators.} The reparameterization from spatial bill positions $(\zeta_j, \psi_j)$ to discrimination parameters $(\alpha_j, \beta_j)$ is many-to-one—entirely different bill locations generate identical parameters—making bills difficult to interpret as spatial entities \citep{Shin2024, nakis2025}. \textcolor{black}{In addition, the model does not include a separate legislator-specific intercept capturing baseline voting propensity; instead, variation across legislators is represented solely through their spatial positions.}

\textcolor{black}{\citet{jeon:2020} introduced the Latent Space Item Response Model (LSIRM) for educational testing, embedding respondents and items in a shared latent space. This symmetric framework applies naturally to roll-call voting data, where legislators and bills form the two modes of a bipartite system. It is also flexible with respect to the geometric operator governing interactions: \citet{jeon:2020} consider both an inner-product specification, analogous to the latent eigenmodel \citep{Hoff:2007}, and a distance-based specification. The former captures directional alignment through $z_i^{\top} w_j$, whereas the latter captures spatial proximity through Euclidean distance. In this paper, we adopt the Euclidean specification.}

Distance-based specifications have been applied to political networks in two contexts, though with different formulations and objectives than ours. \citet{Barbera2015} employed a latent space model to estimate ideal points from Twitter following patterns, treating the data as a bipartite network where users choose whether to follow political actors. Users $i$ and political actors $j$ are embedded in a shared latent space, with following probability modeled as $P(y_{ij}=1) = f(\phi - \gamma \lVert \theta_i - \beta_j \rVert^2)$, where $\lVert \theta_i - \beta_j \rVert^2$ represents squared Euclidean distance. However, \textcolor{black}{as discussed in Section~\ref{sec:theory}, this squared distance formulation does not satisfy the triangle inequality and therefore produces a non-metric spatial representation.}

\textcolor{black}{Taken together, these approaches reveal two limitations in the spatial analysis of roll-call voting. First, in conventional ideal point models bills are treated primarily as scaling instruments with parameters that do not correspond to identifiable spatial locations. Second, distance-based specifications used in political network models typically rely on squared Euclidean distance, which—as discussed in Section~\ref{sec:theory}—does not satisfy the triangle inequality and therefore produces a non-metric representation of ideological space. In the next section, we introduce a Euclidean latent space formulation for roll-call voting that embeds legislators and bills in a shared space using Euclidean distance and allows direct comparison with inner-product-based representations.}

\subsection{Euclidean Distance Specification}

We specify voting probability through Euclidean distance in a bipartite network framework. For legislator $i$ with latent position $z_i \in \mathbb{R}^K$ and bill $j$ with latent position $w_j \in \mathbb{R}^K$:
\begin{equation}
\text{logit}(P(y_{ij} = 1 \mid \theta_i, \beta_j, \gamma, z_i, w_j)) = \theta_i + \beta_j - \gamma \lVert z_i - w_j \rVert,
\label{eq:lsirm}
\end{equation}
where $\theta_i$ captures legislator $i$'s baseline propensity to vote ``Yea'', $\beta_j$ represents bill $j$'s baseline popularity or controversy, and $\gamma > 0$ governs the strength of proximity effects. The Euclidean distance is $\lVert z_i - w_j \rVert = \sqrt{\sum_{k=1}^K (z_{ik} - w_{jk})^2}$. The probability of a ``Yea'' vote increases with legislator propensity $\theta_i$ and bill popularity $\beta_j$, but decreases with spatial distance $\lVert z_i - w_j \rVert$. This formulation directly implements proximity voting \citep{downs1957economic, davis1970expository}: legislators support bills near their ideal points, with support declining monotonically as separation increases.

\textcolor{black}{
For comparison, an inner-product LSIRM can be written as
\begin{equation}
\text{logit}(P(y_{ij} = 1 \mid \theta_i, \beta_j, z_i, w_j)) = \theta_i + \beta_j + z_i^{\top} w_j,
\label{eq:innerlsirm}
\end{equation}
which replaces the distance-based interaction with a bilinear form while preserving the same intercept structure.
}

The baseline parameters $\theta_i$ and $\beta_j$ play distinct roles from \textcolor{black}{those in BIRT. In BIRT, bill positions are not separately identified such that $\alpha_j$ can be interpreted as a baseline level of support, and the model includes no legislator-specific intercept term.} In our specification, $\beta_j$ measures only the baseline tendency for legislators to support bill $j$ averaging over all spatial locations—a pure ``how popular is this bill'' parameter separable from its ideological location $w_j$. Similarly, $\theta_i$ measures legislator $i$'s general tendency to vote ``Yea'' independent of ideological proximity—some legislators vote ``Yea'' frequently (high $\theta_i$) while others vote ``Nay'' frequently (low $\theta_i$) for reasons orthogonal to spatial positioning. This separation between baseline propensities and spatial structure, which is similar to the degree correction in network science literature \citep{fortunato2010community}, clarifies interpretation: if legislator $i$ supports bill $j$, we can determine whether this reflects high baseline propensity ($\theta_i$ or $\beta_j$ large), ideological proximity ($\lVert z_i - w_j \rVert$ small), or both.

The parameter $\gamma$ quantifies the strength of spatial proximity relative to baseline propensities. Large $\gamma$ indicates voting is primarily distance-determined—legislators vote almost exclusively for nearby bills regardless of baseline propensities—while small $\gamma$ suggests the dominance of baseline propensities, a spatial structure playing a secondary role. This provides diagnostic information about legislative organization: chambers with strong, stable party or factional discipline exhibit larger $\gamma$ (tight clustering with sharp spatial boundaries), whereas fragmented or weakly organized chambers exhibit smaller $\gamma$ (diffuse positioning where baseline propensities matter more than ideological location). Our simulation studies demonstrate that $\hat{\gamma}$ responds predictably to known levels of coalition cohesion, increasing when within-group homogeneity strengthens and decreasing when independent voting attenuates the distance signal.

\textcolor{black}{
Inner-product models, including BIRT, measure affinity through directional alignment rather than spatial proximity, so geometric closeness does not necessarily imply a higher tie probability. As noted in the \texttt{latentnet} documentation, actors that are closer in a clustering sense need not have higher expected tie values \citep{Krivitsky:2009}. More generally, as shown by \citet{Hoff:2007}, inner-product formulations are representationally more flexible, which can be advantageous in networks characterized by directional equivalence or heterogeneous connectivity patterns.
}

\textcolor{black}{
By contrast, Euclidean distance models impose a stricter geometric interpretation in which spatial proximity directly governs interaction probability. This yields embeddings that are less sensitive to scale artifacts and limits the influence of unusually prominent units, such as procedural hub bills or omnibus measures. In homophilous settings such as legislative voting, the resulting geometry aligns clustering structure more closely with proximity in the latent space. Consistent with this intuition, \citet{nakis2025} show that Euclidean representations can often achieve comparable structural fit with equal or lower latent dimensionality than inner-product formulations.
}

The bipartite embedding provides three interpretive advantages over conventional methods \textcolor{black}{by locating legislators and bills in a shared latent space. Legislator-to-legislator distances reveal ideological similarity, bill-to-bill distances identify policy clustering, and legislator--bill distances clarify coalition structure through spatial proximity, including the placement of cross-party bridge bills between cooperating factions.} This joint embedding enables interpretation through bill locations as anchors: instead of asking ``what does dimension 2 mean?'', we can observe ``which bills define dimension 2?'' and infer substantive meaning from the specific legislation that separates high from low values.

\subsection{Full Conditionals and Estimation}


We estimate parameters via Markov Chain Monte Carlo (MCMC) sampling with the Gibbs sampling and the metropolis-Hastings algorithm with the following prior specifications:
\begin{equation}\label{eq:prior}
\begin{split}
\theta_i &\sim N(0, \sigma_{\theta}^2), \\
\beta_j &\sim N(0, \sigma_{\beta}^2),\\
\sigma_{\theta}^2 &\sim \mbox{Inv-Gamma}(a_{\sigma},b_{\sigma}),\\
z_i &\sim \mbox{MVN}_K(\mathbf{0}, \mathbf{I}_K), \\
w_j &\sim \mbox{MVN}_K(\mathbf{0}, \mathbf{I}_K), \\
\log{\gamma} &\sim N(\mu_{\gamma}, \sigma_{\gamma}^2).    
\end{split}
\end{equation}

\textcolor{black}{
The overall scale of the latent space is regularized through the prior specification $z_i \sim \mbox{MVN}_K(\mathbf{0}, \mathbf{I}_K)$ and $w_j \sim \mbox{MVN}_K(\mathbf{0}, \mathbf{I}_K)$. Because the likelihood depends on the Euclidean distance $\lVert z_i - w_j \rVert$, it is invariant to joint rescaling of the latent positions and the distance coefficient $\gamma$. The prior therefore constrains the scale of the latent space, allowing $\gamma$ to be interpreted as the strength of distance-based voting relative to this normalization.
}

Then, the joint posterior density of the model is given as
\begin{equation}\label{eq:posterior}
    \begin{split}
    \pi(\boldsymbol{\beta}, \boldsymbol{\theta}, \gamma, {\bf Z}, {\bf W} \mid \mathbf{Y}) &\propto \prod_{i=1}^{N}\prod_{j=1}^{P}P(Y_{ij}=y_{ij} \mid \theta_i, \beta_j, \gamma, z_i, w_j) \\ 
    &\times \pi(\gamma) \prod_{j=1}^{P}\pi(\beta_j) \prod_{i=1}^{N}\pi(\theta_i) \prod_{i=1}^{N}\pi(z_i) \prod_{j=1}^{P}\pi(w_j)
    \end{split}
\end{equation}

The priors and the following posterior distribution can be expressed in equations \ref{eq:prior} and \ref{eq:posterior}. This posterior kernel cannot be expressed with standard distribution, so the exact posterior density cannot be calculated and a Gibbs sampler is used to sample each parameter sequentially from their conditional density. Specifically, because each conditional kernel may not be expressed as a standard distribution form, Metropolis-Hastings within Gibbs sampler is used. Furthermore, for the generalized case with missing data, the imputed value for the non-respondent pair can be sampled with the logistic formula at the start of each Gibbs sampler step. 
\begin{equation*}
\begin{split}
    \pi(\theta_i) &\propto \Bigg[\prod_{i=1}^{N}\prod_{j=1}^{P}P(Y_{ij}=y_{ij} \mid \theta_i, \beta_j, \gamma, z_i, w_j)\Bigg] \times  \big[N_{\theta_i}(0, \sigma_{\theta}^2) \big]\\
    \pi(\beta_j) &\propto \Bigg[\prod_{i=1}^{N}\prod_{j=1}^{P}P(Y_{ij}=y_{ij} \mid \theta_i, \beta_j, \gamma, z_i, w_j)\Bigg] \times  \big[N_{\beta_j}(0, \sigma_{\beta}^2) \big]\\
    \pi(\log \gamma) &\propto \Bigg[\prod_{i=1}^{N}\prod_{j=1}^{P}P(Y_{ij}=y_{ij} \mid \theta_i, \beta_j, \gamma, z_i, w_j)\Bigg] \times  \big[N(\mu_{\gamma}, \sigma_{\gamma}^2)]\\
    \pi(z_i) &\propto \Bigg[\prod_{i=1}^{N}\prod_{j=1}^{P}P(Y_{ij}=y_{ij} \mid \theta_i, \beta_j, \gamma, z_i, w_j)\Bigg] \times  \big[\mbox{MVN}_{K, z_i}(\mathbf{0}, \mathbf{I}_K) \big]
    \end{split}
\end{equation*}

\begin{equation*}
\begin{split}
    \pi(w_j) &\propto \Bigg[\prod_{i=1}^{N}\prod_{j=1}^{P}P(Y_{ij}=y_{ij} \mid \theta_i, \beta_j, \gamma, z_i, w_j)\Bigg] \times  \big[\mbox{MVN}_{K, w_j}(\mathbf{0}, \mathbf{I}_k) \big]\\
    \pi(\sigma_{\theta}) &\propto \mbox{Inv-Gamma}\Bigg(\bigg(\frac{N}{2}+a_{\sigma}\bigg),\frac{1}{2}\sum_{i=1}^{N}{\theta}_i^2 + b_{\sigma}\bigg)
\end{split}
\end{equation*}
Like conventional ideal-point estimators, Euclidean LSIRM inherits rotational, reflective, and translational indeterminacies: for any orthogonal matrix $Q$ and vector $c$, the transformed configuration $(Qz_i + c,\; Qw_j + c)$ produces identical probabilities. Because we do not fix the positions of any legislators or bills, the absolute orientation and location of the latent space are not identified. \textcolor{black}{Specifically, after obtaining posterior samples from the MCMC algorithm, we apply a Procrustes transformation \citep{gower_generalized_1975} to align each sampled configuration to a reference configuration selected as the draw with the highest posterior density \citep{Friel:2016}. This post-processing removes rotational, reflective, and translational indeterminacies while preserving pairwise distances among legislators and bills.}\footnote{Recently, \citet{Shin2024} proposed an $\ell_1$-norm based Bayesian IRT model, which transforms the infinite rotational invariance of conventional 
methods into a more tractable signed perpendicular rotational invariance.}

\section{Simulation Studies}\label{sec:simulation}

We designed three simulations to evaluate Euclidean LSIRM against \textcolor{black}{two alternative formulations: BIRT, a widely used ideal-point model, and an inner-product LSIRM specification in which interactions between legislators and bills are modeled through the inner product of latent positions. To isolate the role of latent-space geometry, we constructed a comparison between Euclidean and inner-product LSIRM specifications that share the same baseline parameters, bill representation, scaling structure, and identification procedure, differing only in the geometric form of interaction. Because BIRT also represents legislator--bill interactions through an inner-product structure between legislator ideal points and bill parameters, these two approaches provide useful contrasts to the Euclidean-distance formulation of LSIRM. The simulations progressively increase in structural complexity and address three questions:}
\begin{enumerate}
    \item Cohesion Gradient: Does the proximity parameter $\gamma$ respond appropriately to known levels of group cohesion?
    \item Cluster Recovery: Can the model recover multiple independent clusters in low-dimensional space?
    \item Party--Faction Structure: Can the model distinguish factions within parties and cross-party coalitions?
\end{enumerate}

All simulations used 100 replicated datasets. MCMC chains ran for 30,000 iterations with 5,000 burn-in and thinning factor 5. We evaluate performance using silhouette coefficients, and visual inspection of latent embeddings. \textcolor{black}{ For visualization, we display representative replications whose silhouette coefficients across the three models are jointly closest to their median values.}

\textcolor{black}{In the main text, we focus on the party--faction simulation, which is most directly connected to the paper's goal of recovering within-party factions and cross-party coalitions. Additional results for the cohesion-gradient and cluster-recovery designs are reported in Supplementary Materials (Section 1.1 and 1.2), while predictive accuracy and log-likelihood summaries across settings are reported in Supplementary Materials (Section 1.4).} 

\textcolor{black}{The party--faction simulation considers} two parties with internal factions, plus cross-party coalitions. The baseline structure consists of two parties, each containing two or three factions. We generated three types of bills: \emph{partisan bills} supported by one party against the other, \emph{faction bills} supported by a single faction, and \emph{bridge bills} supported by factions from opposing parties.

\begin{table}[htbp]
\centering
\small
\begin{tabular}{p{3.3cm} p{4.3cm} p{5.3cm}}
\toprule
\textbf{Scenario} & \textbf{Manipulation} & \textbf{Key Test} \\
\midrule
(i) Agenda Sweep & Vary partisan vs.\ faction bill ratio & Faction recovery as partisan signal weakens \\
(ii) Cross-Party Coalitions & Add bridge bills & Representation of overlapping coalitions \\
\bottomrule
\end{tabular}
\caption{Party--faction scenarios examined in the main text. Additional noise-sweep results are reported in Supplementary Materials (Section 1.3.1).}
\label{tab:sim3-overview}
\end{table}

\subsection{Agenda Sweep}

Each party contains three factions (180 legislators total). We varied the share of partisan bills from 30\% to 50\%, with the remainder being faction-specific bills.

\begin{figure}[htb]
\centering
\begin{subfigure}[t]{0.3\linewidth}
    \centering
    \includegraphics[width=\linewidth]{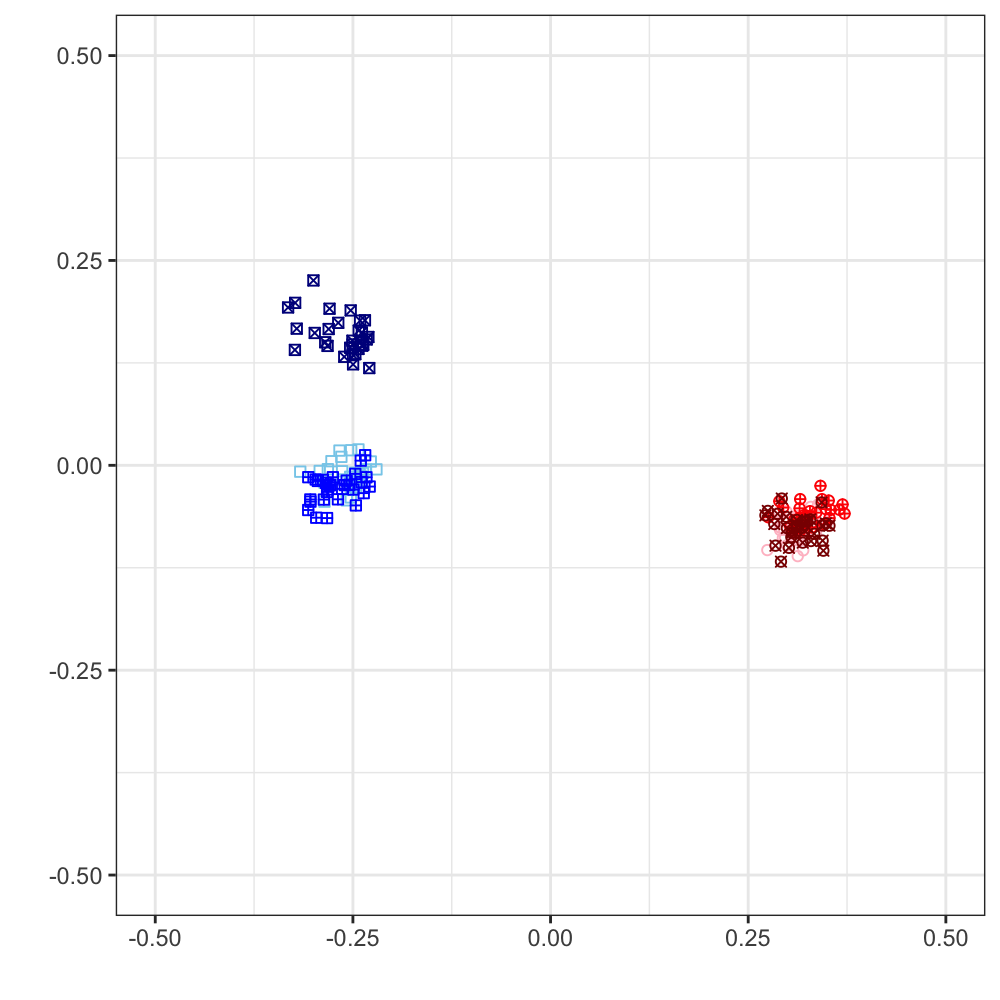}
    \caption{BIRT (50\% partisan)}
\end{subfigure}\hfill
\begin{subfigure}[t]{0.3\linewidth}
    \centering
    \includegraphics[width=\linewidth]{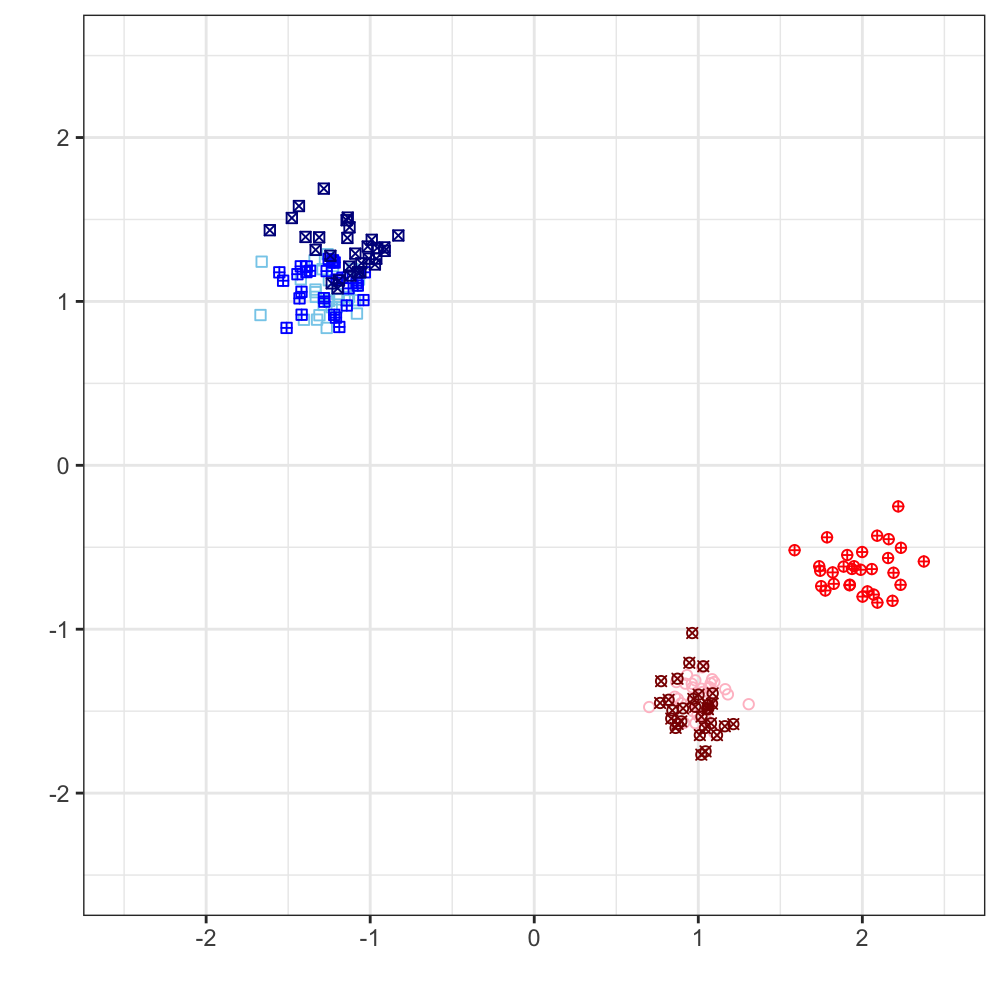}
    \caption{Inner-product LSIRM (50\% partisan)}
\end{subfigure}\hfill
\begin{subfigure}[t]{0.3\linewidth}
    \centering
    \includegraphics[width=\linewidth]{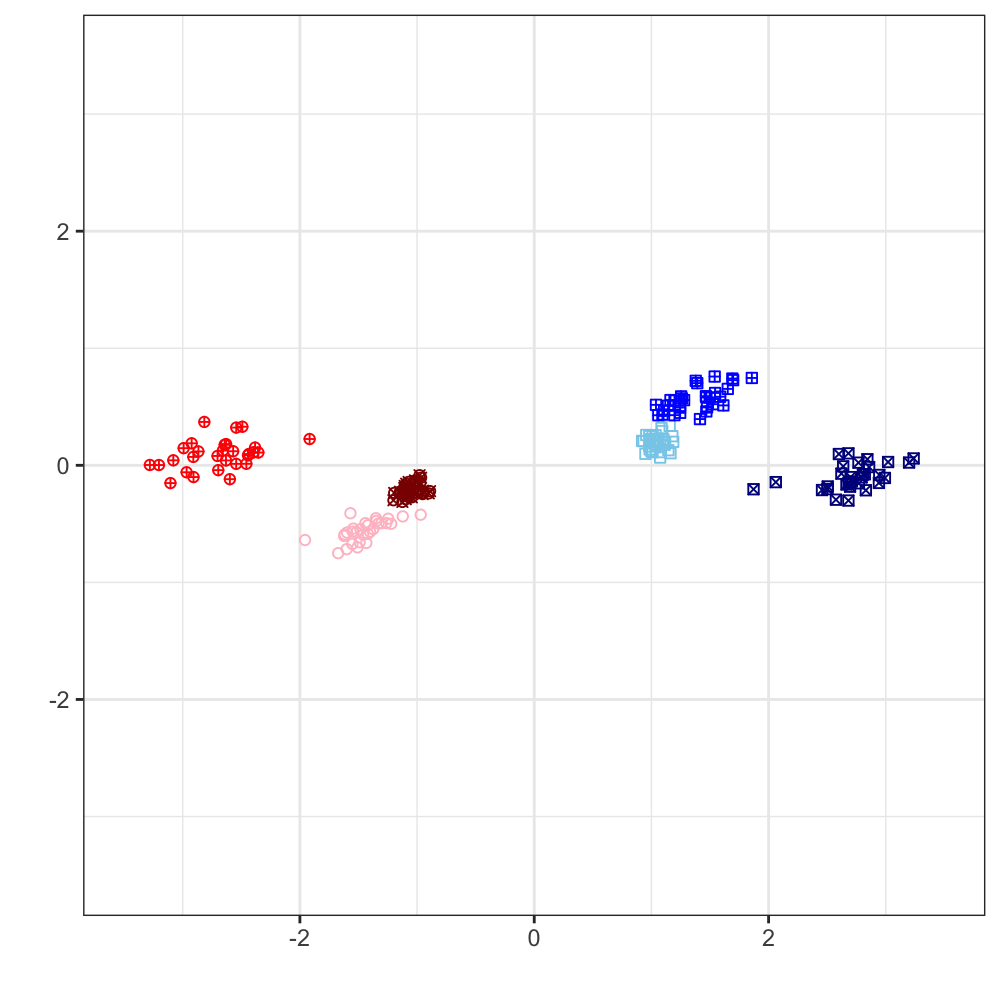}
    \caption{Euclidean LSIRM (50\% partisan)}
\end{subfigure}

\vspace{0.3em}

\begin{subfigure}[t]{0.3\linewidth}
    \centering
    \includegraphics[width=\linewidth]{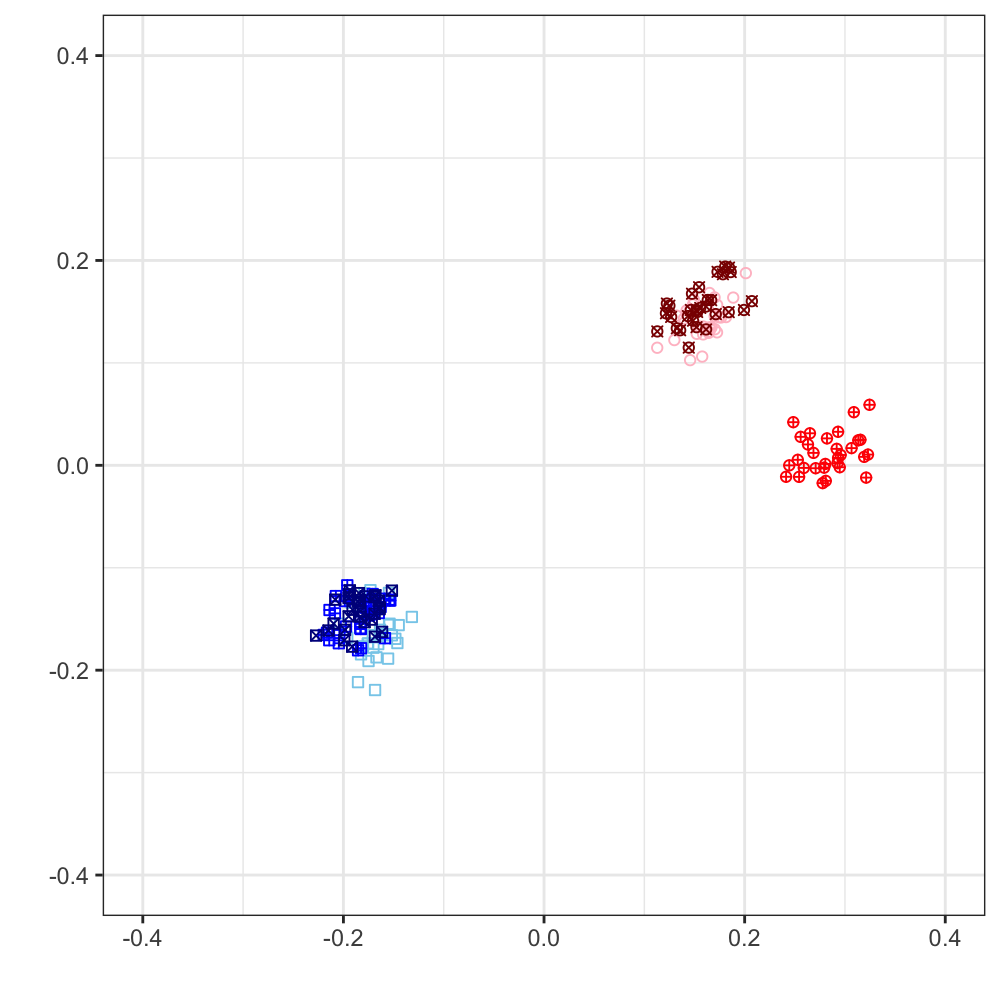}
    \caption{BIRT (30\% partisan)}
\end{subfigure}\hfill
\begin{subfigure}[t]{0.3\linewidth}
    \centering
    \includegraphics[width=\linewidth]{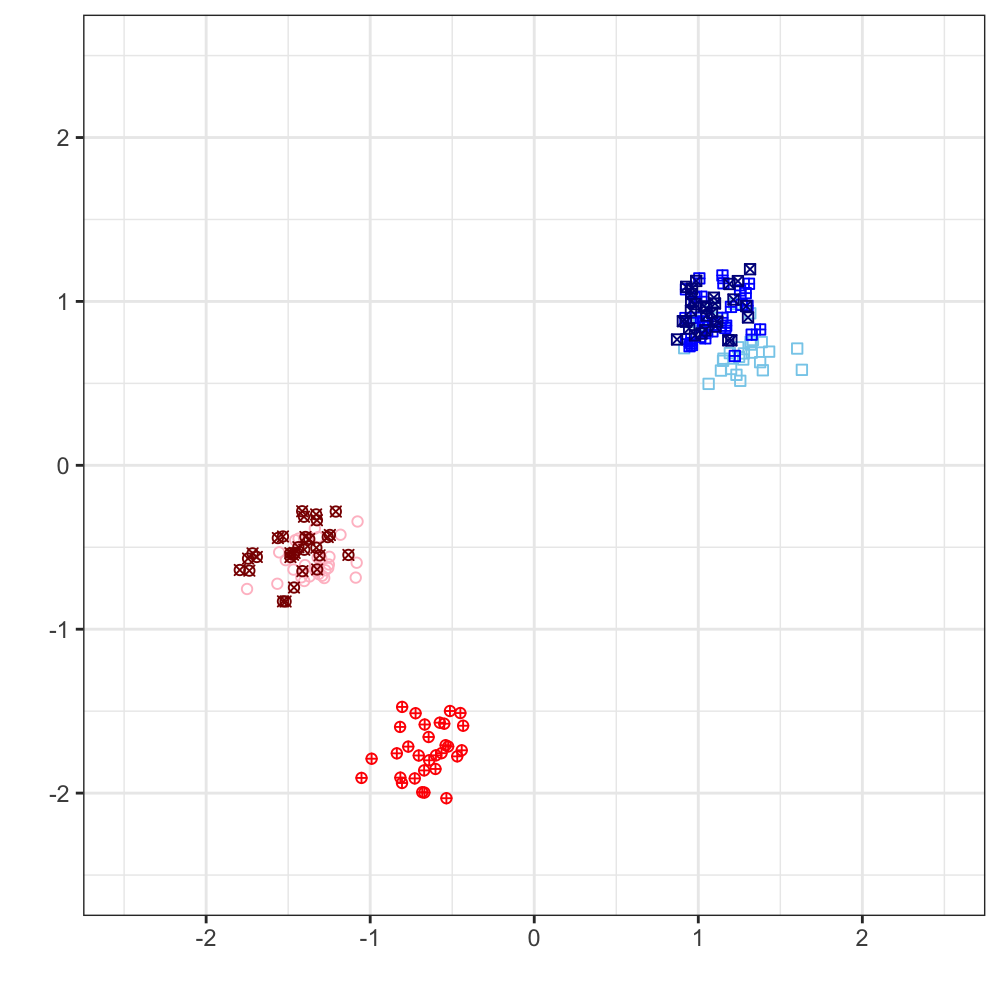}
    \caption{Inner-product LSIRM (30\% partisan)}
\end{subfigure}\hfill
\begin{subfigure}[t]{0.3\linewidth}
    \centering
    \includegraphics[width=\linewidth]{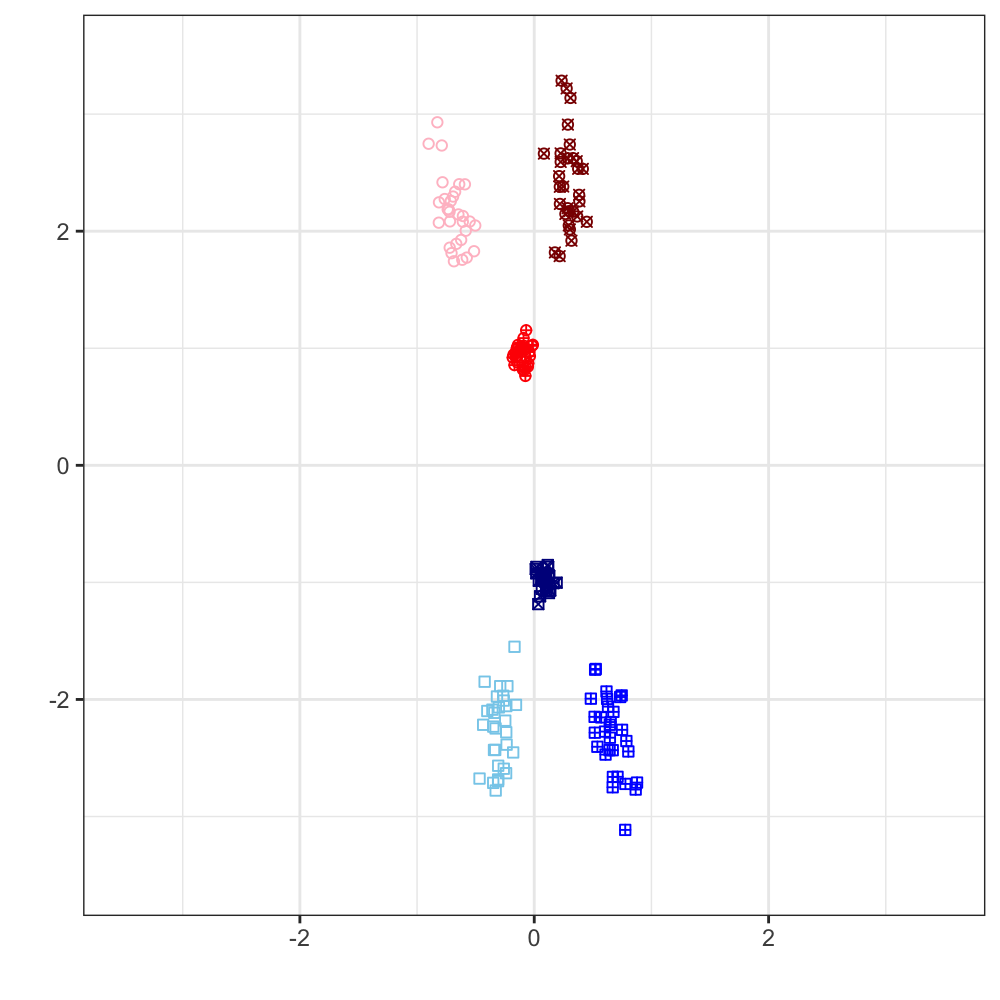}
    \caption{Euclidean LSIRM (30\% partisan)}
\end{subfigure}
\caption{\textcolor{black}{Faction recovery under shifting legislative agendas. Colors and markers denote 
six true factions: two parties, each containing three internal factions (180 legislators 
total, 30 per faction). Top row: 50\% partisan bills; bottom row: 30\% partisan bills. 
Euclidean LSIRM recovers all six factions as distinct clusters across both conditions, 
whereas BIRT and inner-product LSIRM collapse within-party factions into roughly three 
visible groups, failing to distinguish factional structure from party membership.}}
\label{fig:sim3-agenda}
\end{figure}



\textcolor{black}{At both agenda compositions, BIRT and the inner-product LSIRM largely collapse factional structure, merging within-party factions into roughly three visible groups in the representative embedding (Figure~\ref{fig:sim3-agenda}). Euclidean LSIRM, by contrast, separates all six factions into distinct clusters. As the share of partisan bills falls from 50\% to 30\%, factional signals strengthen and the performance gap widens: Euclidean LSIRM's mean silhouette increases from $0.624$ ($SD = 0.081$) to $0.723$ ($SD = 0.016$), while BIRT remains low ($0.164$ to $0.172$) and the inner-product LSIRM shows similarly weak separation ($0.188$ to $0.190$).}

\subsection{Cross-Party Coalitions}
We introduced unequal faction sizes (70 vs.\ 10 legislators per faction) and two types 
of bridge bills: \emph{Bridge-A} supported by majority factions from both parties (L1 
and C1), and \emph{Bridge-B} supported by minority factions (L2 and C2). This design 
tests whether the model can represent overlapping coalitions that transcend party lines. 
Additional configurations and extended results, including simulations incorporating 
faction bills, appear in the \textcolor{black}{Supplementary Materials (Section 1.3.2)}.
\begin{figure}[htbp]
\centering
\begin{subfigure}[t]{0.31\linewidth}
    \centering
    \includegraphics[width=\linewidth]{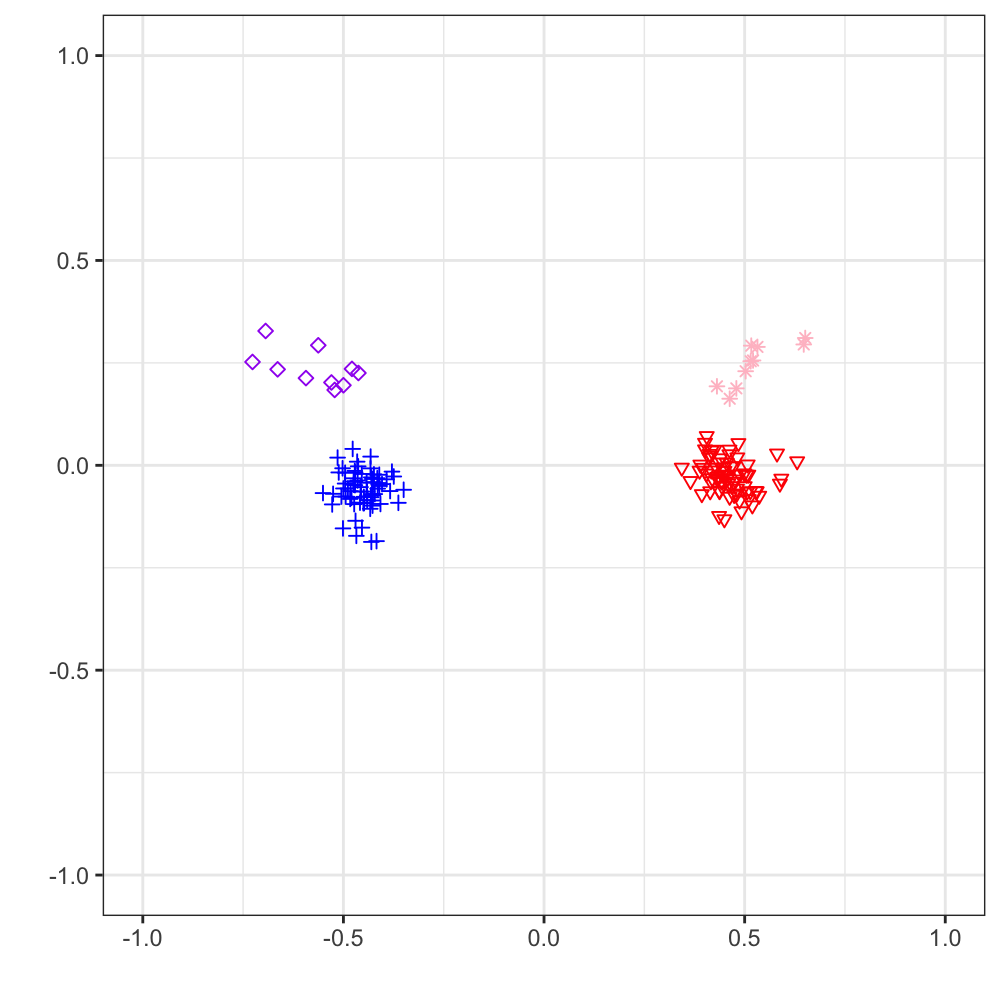}
    \caption{BIRT}
\end{subfigure}\hfill
\begin{subfigure}[t]{0.31\linewidth}
    \centering
    \includegraphics[width=\linewidth]{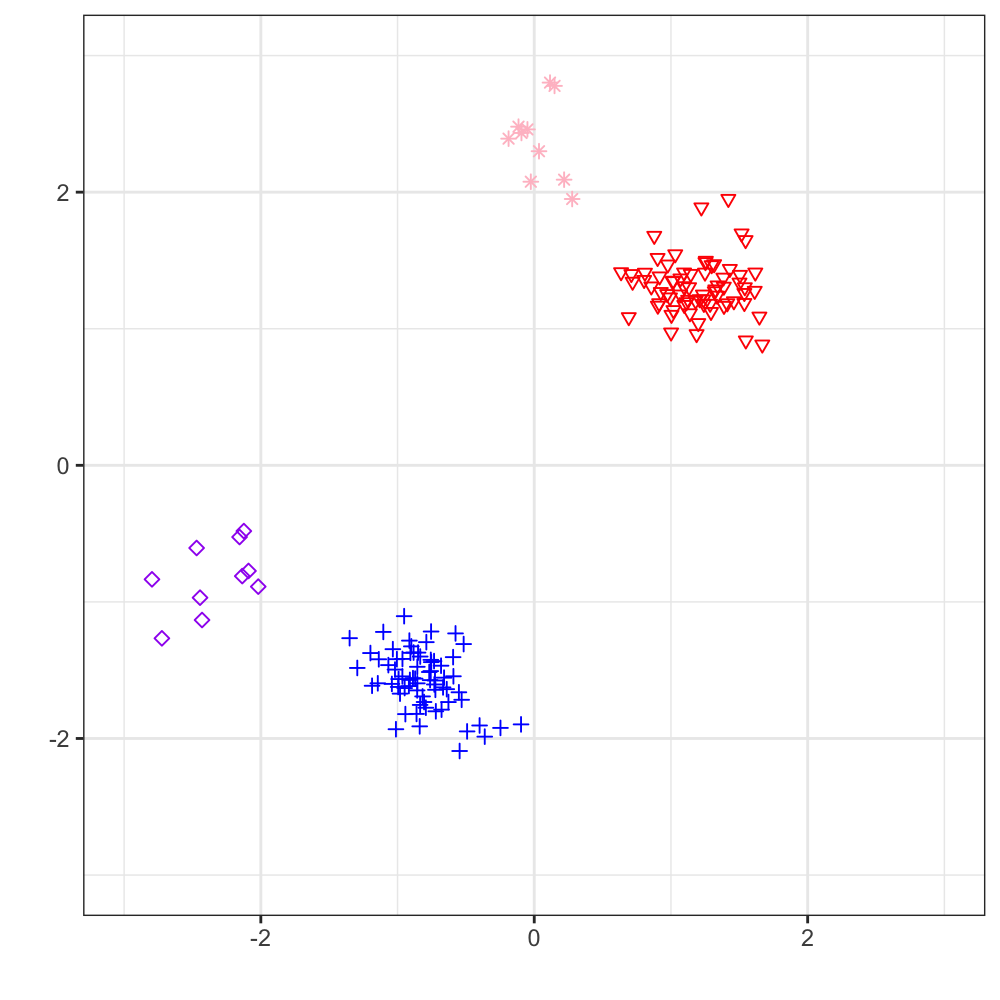}
    \caption{Inner-product LSIRM}
\end{subfigure}\hfill
\begin{subfigure}[t]{0.31\linewidth}
    \centering
    \includegraphics[width=\linewidth]{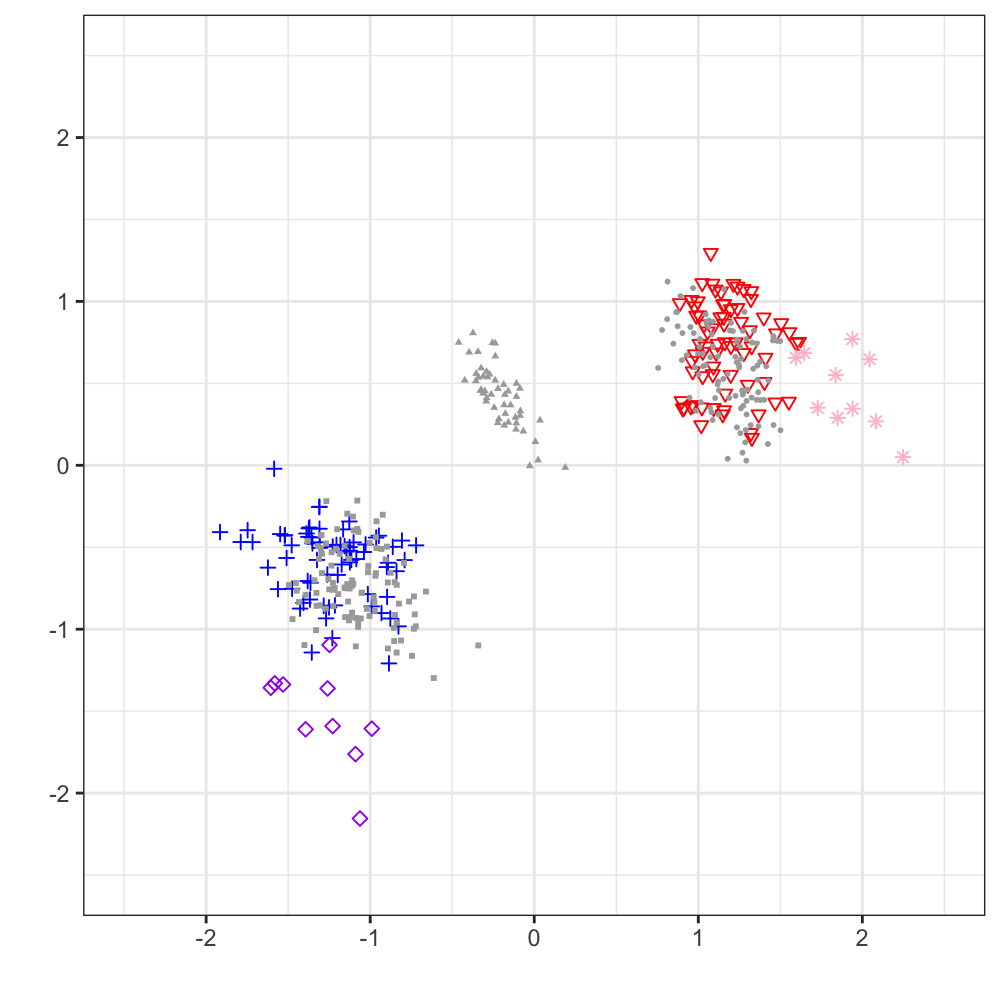}
    \caption{Euclidean LSIRM}
\end{subfigure}
\caption{\textcolor{black}{Cross-party majority coalition recovery. Two parties (Liberal: blue/purple; 
Conservative: red/pink) each split into a majority faction (L1, C1: 70 legislators) 
and a minority faction (L2, C2: 10 legislators). Bridge-A bills (grey triangles) are 
supported jointly by the two majority factions across party lines. Legislators are 
colored by party and shaped by faction (L1: blue squares, L2: purple triangles, C1: 
red circles, C2: pink crosses). Euclidean LSIRM correctly positions allied majority 
factions near the inter-party boundary with Bridge-A bills between them; BIRT and 
inner-product LSIRM fail to capture this cross-party coalition structure.}}
\label{fig:crossparty-majority}
\end{figure}

\textcolor{black}{In the majority consensus scenario (Figure~\ref{fig:crossparty-majority}), 
Euclidean LSIRM correctly draws the allied majority factions closer together near the 
inter-party boundary, with Bridge-A bills positioned between them. Both BIRT and the 
inner-product LSIRM fail to represent this structure---distances between allied 
cross-party legislators are no shorter than distances between ideological opponents, 
obscuring the coalition structure in the latent space.}

\begin{figure}[htbp]
\centering
\begin{subfigure}[t]{0.31\linewidth}
    \centering
    \includegraphics[width=\linewidth]{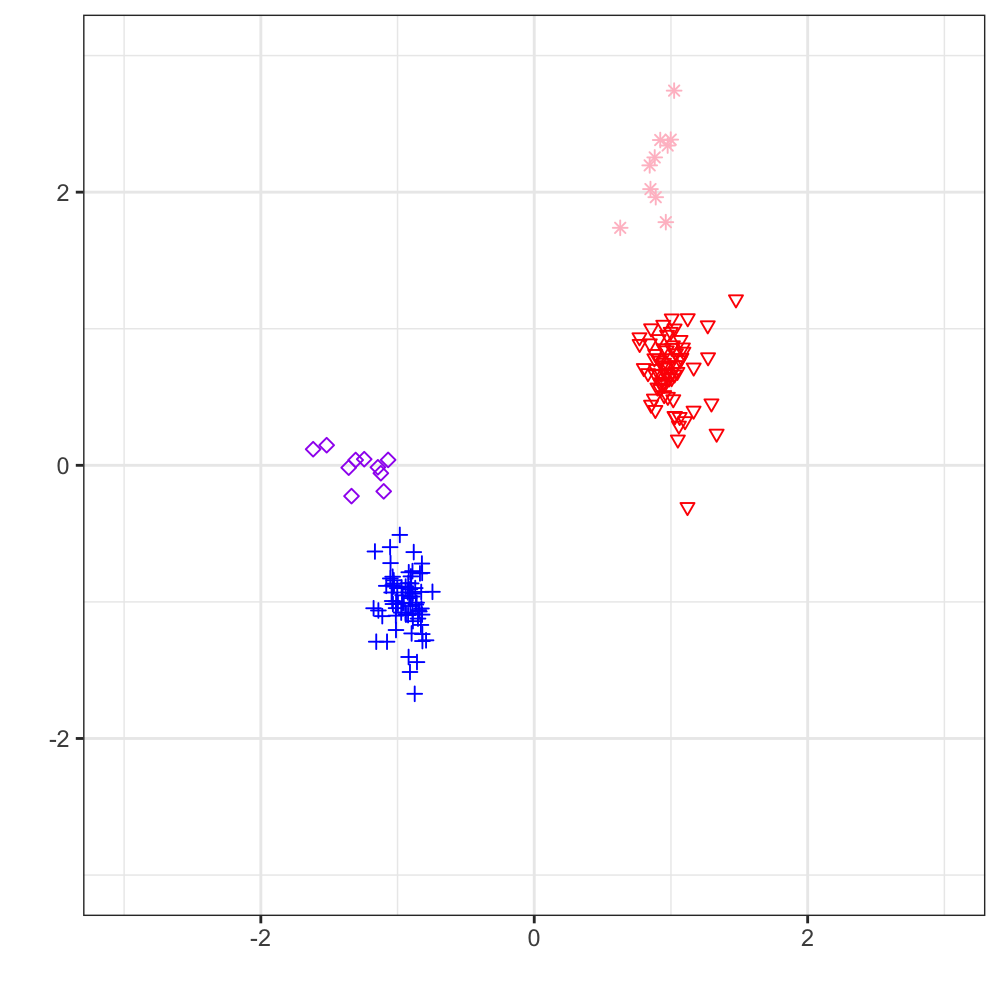}
    \caption{BIRT}
\end{subfigure}\hfill
\begin{subfigure}[t]{0.31\linewidth}
    \centering
    \includegraphics[width=\linewidth]{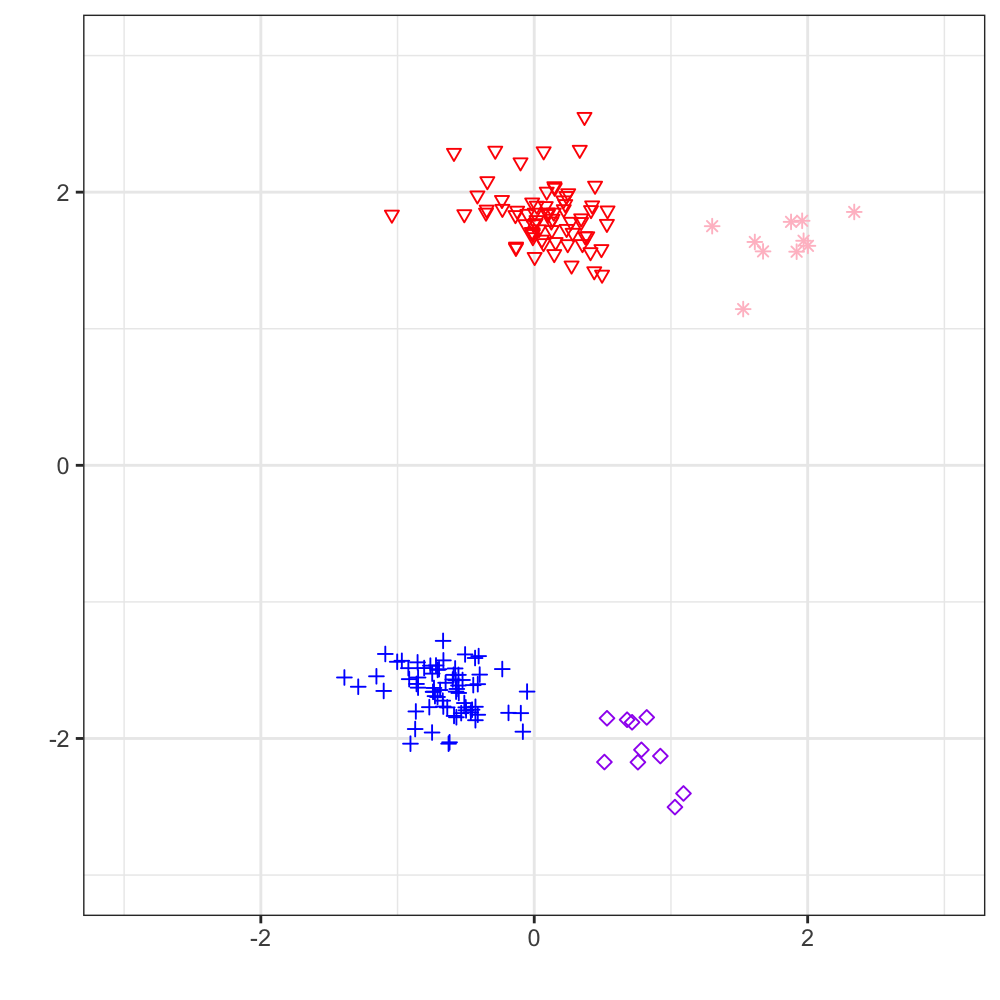}
    \caption{Inner-product LSIRM}
\end{subfigure}\hfill
\begin{subfigure}[t]{0.31\linewidth}
    \centering
    \includegraphics[width=\linewidth]{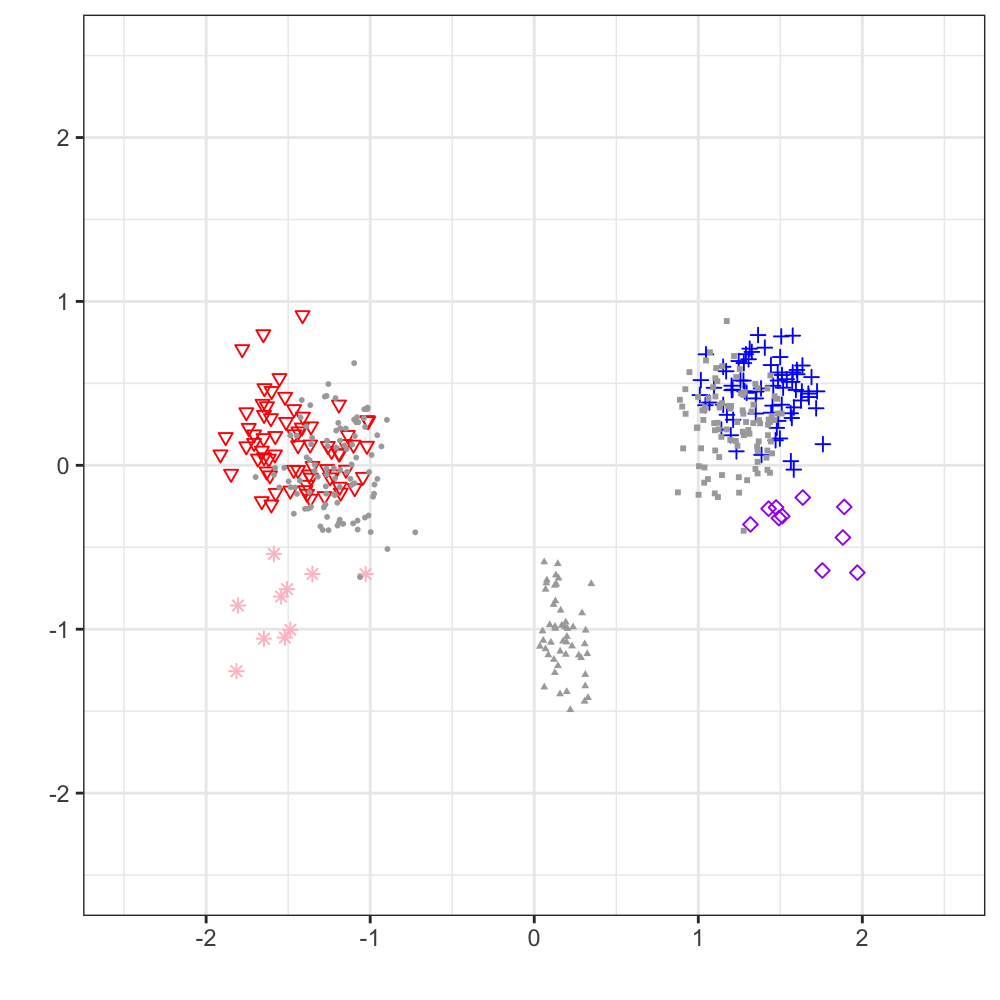}
    \caption{Euclidean LSIRM}
\end{subfigure}
\caption{\textcolor{black}{Cross-party minority coalition recovery: ends-against-the-middle scenario. 
Bridge-B bills (grey triangles) are supported jointly by the two minority factions 
(L2, C2: 10 legislators each) across party lines, while majority factions (L1, C1: 
70 legislators each) oppose. Legislators are colored by party and shaped by faction 
(L1: blue squares, L2: purple triangles, C1: red circles, C2: pink crosses). Euclidean 
LSIRM draws the minority factions toward each other with Bridge-B bills positioned 
near this pairing, preserving interpretable proximity between legislators and the bills 
structuring their coalition. BIRT and inner-product LSIRM instead stretch legislators 
along a pronounced secondary axis, obscuring the minority coalition.}}
\label{fig:crossparty-minority}
\end{figure}

\textcolor{black}{In the ends-against-the-middle scenario \citep{Duck-Mayr:2023}, the 
two inner-product specifications---BIRT and the inner-product LSIRM---produce embeddings 
that stretch legislators along a pronounced secondary axis (Figure~\ref{fig:crossparty-minority}). 
This elongated geometry spreads factions along a single direction in the latent space.}

\textcolor{black}{The Euclidean LSIRM instead organizes legislators according to spatial 
proximity. As a result, the two minority factions are drawn toward one another while the 
majority blocs remain separated. This configuration places Bridge-B bills between the 
opposing minority factions and preserves interpretable distances between legislators and 
the bills that structure their voting behavior.}



\section{Application: 118th U.S. House of Representatives}\label{sec:application}

\subsection{Data and Context}

We apply Euclidean LSIRM to roll-call votes from the 118th U.S. House, obtained from Voteview \citep{voteview2025}. Following standard practice, we excluded lopsided votes (overwhelming majorities) that provide little ideological information, yielding 1,225 informative roll-calls across 451 unique legislators (accounting for vacancies and replacements). This chamber provides a stringent test: extreme partisan polarization (the most conservative Democrat remains left of the most moderate Republican) yet well-documented intra-party factions—the progressive ``Squad'' (approximately 8 members) among Democrats and the House Freedom Caucus (approximately 22 members) among Republicans. These factions, though numerically small, exert disproportionate influence on leadership contests and agenda setting, making their accurate identification substantively important. \textcolor{black}{We ran five independent MCMC chains and assessed convergence using overlaid trace plots and scale reduction diagnostics (see Supplementary Materials, Section 2.1), which indicate satisfactory mixing and stationarity. As the latent space is invariant to rigid transformations, coordinates are not directly comparable across chains; visualizations therefore use posterior means from the first converged chain.}

\subsection{Legislator Positions and Faction Recovery}
\begin{figure}[htbp]
    \centering
    \begin{minipage}{0.47\linewidth}
        \centering
        \includegraphics[width=\linewidth]{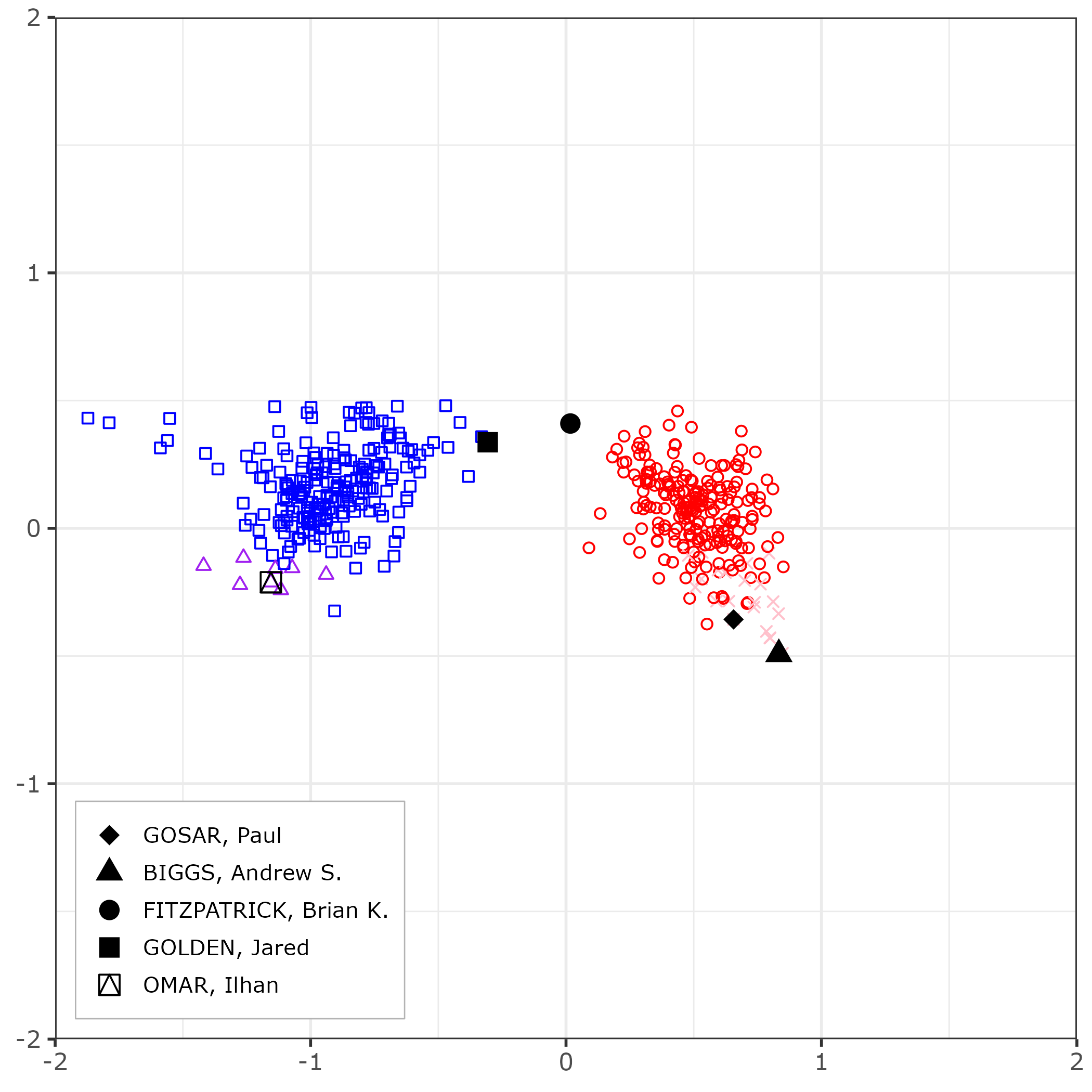}
        \text{(a) Bayesian IRT}
    \end{minipage}
    \begin{minipage}{0.47\linewidth}
        \centering
        \includegraphics[width=\linewidth]{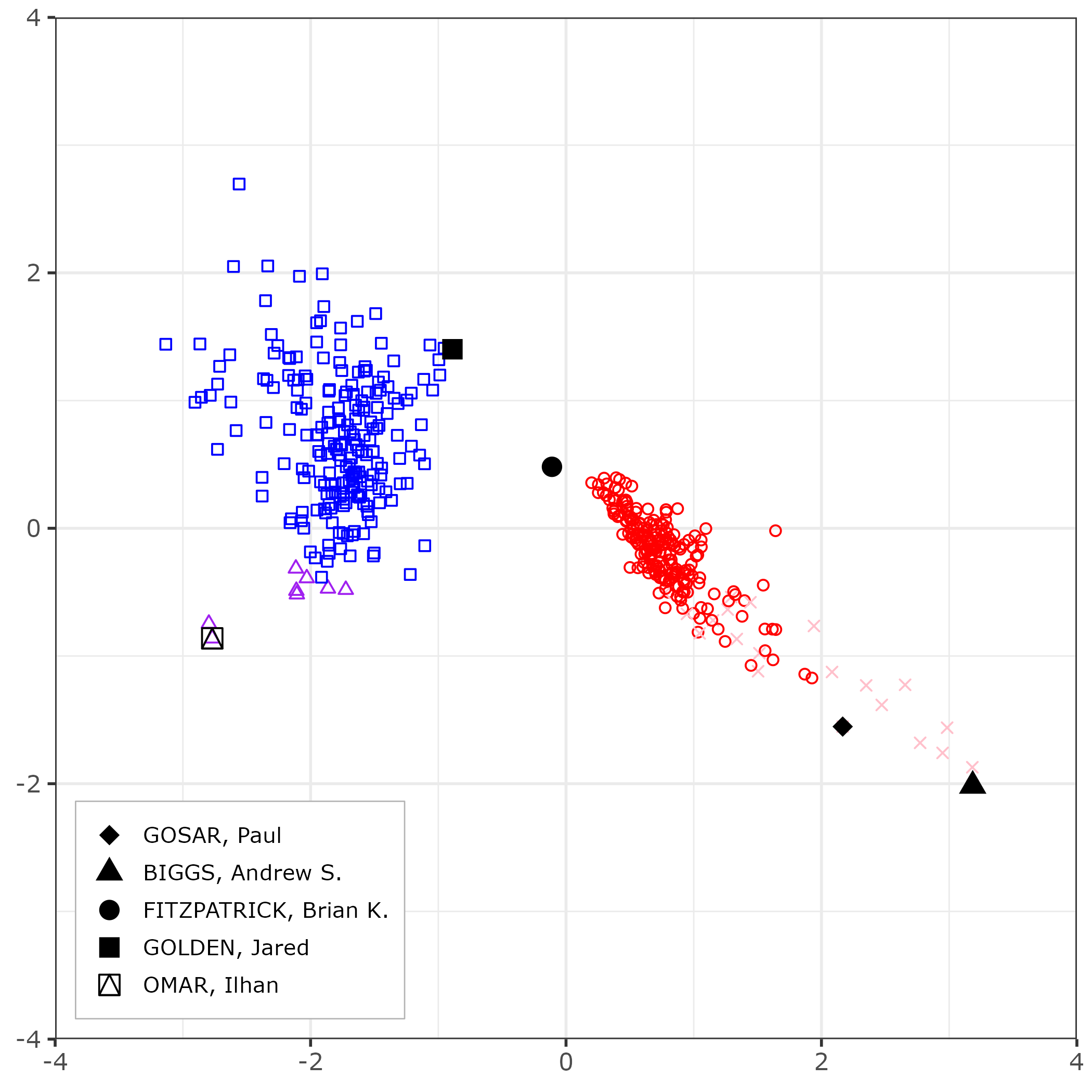}
        \text{(b) Euclidean LSIRM}  
    \end{minipage}
\caption{
\textcolor{black}{Two-dimensional legislator ideal point estimates for the 118th U.S. House.} Legislators are plotted by party and factional affiliation: Mainstream Democrats (blue squares), ``Squad'' Democrats (purple triangles), Mainstream Republicans (red circles), and Freedom Caucus Republicans (pink crosses).
}
\label{fig:5-2-1}
\end{figure}

Figure \ref{fig:5-2-1} displays two-dimensional embeddings from Euclidean LSIRM and BIRT. Both methods recover clear party separation on the primary (horizontal) dimension, with Democrats left and Republicans right. However, the secondary dimension reveals striking differences. Euclidean LSIRM differentiates factions: Freedom Caucus members (e.g., Andy Biggs, Paul Gosar) occupy far-right positions with distinctly negative second-dimension coordinates, separating them from mainstream Republicans, while Squad members (e.g., Ilhan Omar, Alexandria Ocasio-Cortez) appear below with slightly negative values, distinguishing them from other Democrats. The second dimension has substantial variance (SD = \textcolor{black}{0.69}) in Euclidean LSIRM, indicating genuine factional structure. BIRT places faction members simply at extremes of the left-right continuum without meaningful secondary differentiation (SD of dimension 2 = 0.18 versus \textcolor{black}{0.75} for dimension 1), effectively collapsing multi-dimensional coalition structure into one-dimensional ordering.

\begin{table}[!b]
\centering
\begin{tabular}{lcc}
\toprule
& LSIRM & BIRT \\
\midrule
\multicolumn{3}{l}{\textit{In-sample}} \\
Accuracy (ACC) & 0.943 & \textcolor{black}{0.942} \\
APRE & 0.841 & \textcolor{black}{0.839} \\
Mean LPPD & -0.154 & \textcolor{black}{-0.156} \\
\midrule
\multicolumn{3}{l}{\textit{Out-of-sample}} \\
Mean LPPD & -0.263 & \textcolor{black}{-0.165} \\
Brier score & 0.075 & \textcolor{black}{0.046} \\
\bottomrule
\end{tabular}
\caption{\textcolor{black}{In-sample and out-of-sample predictive performance for LSIRM and BIRT ($K=2$). Out-of-sample results are based on a 20\% entry-wise holdout.}}
\label{tab:performance}
\end{table}

We evaluate predictive performance using classification accuracy (ACC), 
aggregate proportional reduction in error (APRE), and mean log posterior predictive 
density (LPPD). As shown in Table~\ref{tab:performance}, \textcolor{black}{Euclidean LSIRM and BIRT exhibit very similar in-sample predictive performance. LSIRM is slightly better on in-sample ACC, APRE, and mean LPPD, but the differences are small.} Out-of-sample performance is assessed using a 20\% entry-wise holdout split. \textcolor{black}{The held-out results also show that both models provide strong probabilistic predictions relative to an uninformative baseline.} The Brier score measures the mean squared difference between 
predicted probabilities and actual binary outcomes, ranging from 0 to 1 with lower 
values indicating better probabilistic calibration; the uninformative baseline of 
always predicting $\hat{p} = 0.5$ yields a score of 0.25. \textcolor{black}{Both models substantially improve upon this baseline, with Brier scores well below 0.25. These validation results indicate that Euclidean LSIRM remains predictively competitive with BIRT.}

\textcolor{black}{Beyond predictive performance, the Euclidean specification provides a useful representation of coalition structure through its joint metric embedding of legislators and bills.}  This has broader implications for ideal point estimation: 
legislatures where intra-party factionalism is consequential---such as those with 
multi-party systems, coalition governments, or cross-cutting cleavages that do not map 
cleanly onto a single left--right axis---are precisely the settings where a non-metric 
distance formulation is most likely to obscure latent structure. In such contexts, 
recovering faction-level ideal points accurately is not only a methodological concern 
but a substantive one, as mismeasured positions can distort inferences about coalition 
formation, agenda influence, and legislative bargaining.

\textcolor{black}{Predictive gains from increasing dimensionality diminish beyond $K=2$ for Euclidean LSIRM (see Supplementary Materials Table~B2). Because improvements beyond two dimensions are modest and the substantive coalition structure is already clearly recovered in two dimensions, we adopt $K=2$ for interpretation in the main analysis.}

\subsection{Bill Embeddings as Interpretive Anchors}

\begin{figure}[htbp]
    \centering
    \begin{minipage}{0.47\linewidth}
        \centering
        \includegraphics[width=\linewidth]{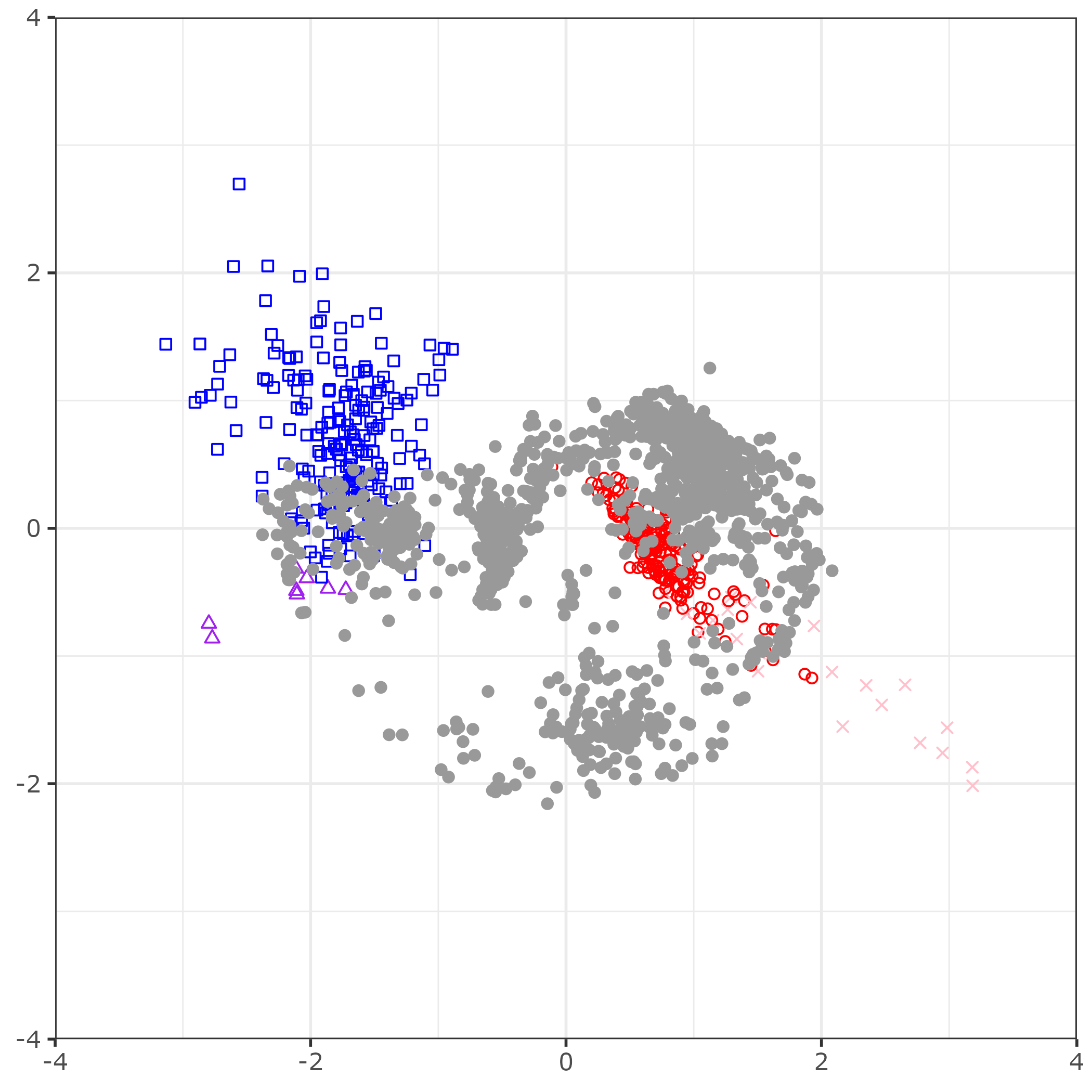}
        \text{(a) Interaction Map}
    \end{minipage}
    \begin{minipage}{0.47\linewidth}
        \centering
        \includegraphics[width=\linewidth]{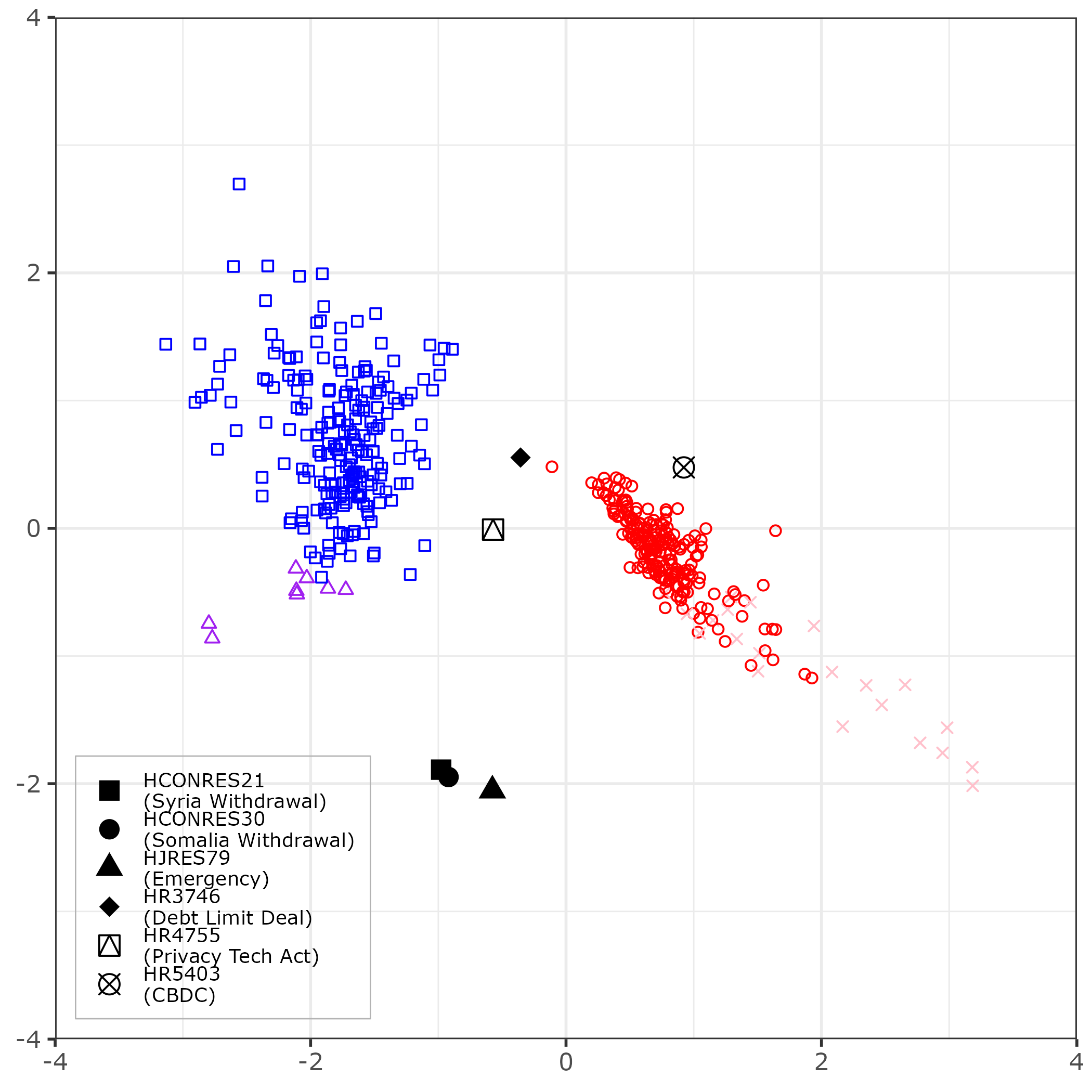}
        \text{(b) Specific Bills}  
    \end{minipage}
\caption{
\textcolor{black}{Two-dimensional legislator--bill ideal point estimates for the 118th U.S. House.} Legislators are plotted by party and factional affiliation: Mainstream Democrats (blue squares), ``Squad'' Democrats (purple triangles), Mainstream Republicans (red circles), and Freedom Caucus Republicans (pink crosses). Bills are represented by gray markers.
}
\label{fig:5-2-2}
\end{figure}

A key advantage of the bipartite network approach is that bill locations serve as interpretive anchors clarifying the substantive meaning of latent dimensions. Figure \ref{fig:5-2-2} displays the full legislator-bill embedding. Partisan bills cluster near their respective party centroids: Republican-supported bills (e.g., tax cuts, defense spending) locate in the far-right region near mainstream Republicans, while Democratic-supported bills (e.g., social programs, environmental regulation) cluster in the left region near mainstream Democrats. This validates that the first dimension captures the familiar liberal-conservative divide.

More revealing are bills that deviate from party centroids. War powers resolutions H.Con.Res. 21 (Syria withdrawal), H.Con.Res. 30 (Somalia withdrawal) and H.J.Res. 79 (terminating national emergency) position in the lower-left quadrant between Squad and Freedom Caucus locations (coordinates approximately $x \approx -0.5$, $y \approx -2$). These resolutions received support primarily from these two factions—progressive Democrats and libertarian Republicans—while being opposed by party establishments. Similarly, surveillance and privacy bills including H.R. 4755 (Privacy Enhancing Technology Research Act), H.R. 5403 (CBDC Anti-Surveillance State Act) locate far from the two extremist factions. These bill positions \textcolor{black}{suggest} the substantive meaning of the second dimension: it \textcolor{black}{reflects} an establishment-outsider cleavage where ideological extremes from both parties converge against centrists on issues of foreign military intervention and government surveillance—the ``horseshoe'' pattern where far-left and far-right align against the center \citep{Duck-Mayr:2023}.

The debt-limit deal H.R. 3746 positions near the center in LSIRM's embedding ($x \approx -0.4$, $y \approx 0.5$), reflecting its passage through a Democratic-anchored coalition with substantial Republican support—a bipartisan compromise occupying ideological middle ground. In contrast, BIRT cannot provide this interpretation because bills are not embedded spatially. The discrimination parameter $\beta_j$ measures directional alignment but does not locate bills in the ideological space, precluding direct visualization of which bills define which dimensions. Without bill embeddings, conventional methods cannot directly show where such compromise legislation locates relative to partisan bills and extremist-coalition bills, obscuring the multi-dimensional coalition structure that characterizes contemporary congressional politics. Additional analyses in the \textcolor{black}{Supplementary Materials (Section 2.4)} visualize bill locations together with within-faction Yea ratios, providing a finer-grained view of how factional support patterns align with Euclidean LSIRM’s spatial geometry.

\section{Discussion}\label{sec:conclusion}

Conventional ideal point estimation methods employ utility functions that violate the triangle inequality, \textcolor{black}{complicating} meaningful interpretation of distance magnitudes and ratios. \textcolor{black}{We show that Euclidean LSIRM provides a more interpretable representation for clustering and coalition analysis.} In controlled simulations with known four-coalition structures, NOMINATE and BIRT compress factions into party blobs (silhouette coefficients: 0.778), while Euclidean LSIRM recovers the true structure (0.861). \textcolor{black}{Application to the 118th House illustrates how joint bill embeddings clarify legislative coalition structure by placing war powers and surveillance bills between Squad and Freedom Caucus members, revealing an establishment-outsider dimension.}

\textcolor{black}{Beyond restoring metric validity, the Euclidean formulation offers a parsimonious and interpretable representation for homophilous roll-call settings. Prior work shows that Euclidean embeddings can often match the structural fit of inner-product formulations with equal or lower latent dimensionality \citep{nakis2025}, while the dependence on pairwise distance rather than vector magnitude reduces sensitivity to scale artifacts or unusually influential units. Most importantly, metric coherence ensures that pairwise similarities satisfy the triangle inequality, supporting clustering, dispersion summaries, and other distance-based interpretations in a way that non-metric formulations cannot.}

These combined properties establish Euclidean LSIRM as particularly well suited for bipartite roll-call analysis when the goal extends beyond ordinal scaling. The bipartite network framework, by treating legislators and bills symmetrically in a shared metric space, enables bill locations to serve as interpretive anchors that define dimensional meaning and clarify which issues generate cross-cutting cleavages. This restores consistency between spatial voting theory, which assumes proximity preferences over metric distances, and empirical estimation, which should produce estimates that support distance-based inference.

Three extensions merit development. First, dynamic specifications tracking coalition reconfiguration across sessions could model legislator positions as evolving while maintaining metric structure within periods, enabling analysis of how factional boundaries shift over time. Second, incorporating bill metadata (i.e., texts, issue codes, sponsors) through hierarchical priors could sharpen interpretation by partially pooling bill locations within policy domains while permitting cross-issue variation. Third, applications to multiparty systems could further exploit Euclidean distance's dimensional efficiency, potentially enabling more parsimonious representations of complex coalition structures in fragmented legislatures where conventional methods struggle with high dimensionality. \textcolor{black}{A related extension would apply the framework to historical congressional sessions, such as those of the 1930s or 1960s, where cross-party factional structures are well documented and provide a richer setting for evaluating clustering-based representations. } Each extension would require addressing identification and missing data issues particular to its context, but the fundamental advantages—dimensional efficiency, balanced influence, and metric coherence—apply broadly across institutional settings where researchers seek to identify voting coalitions rather than merely scale legislators ordinally.

\paragraph{Acknowledgments}
The authors thank the editor, associate editor, and reviewers for their constructive comments. Kwangok Seo and Johan Lim provided valuable comments for the earlier version of this paper. Co-correspondence should be addressed to Jong Hee Park, Department of Political Science and International Relations, Seoul National University, Seoul, Republic of Korea, E-mail: jongheepark@snu.ac.kr, and Ick Hoon Jin, Department of Applied Statistics, Department of Statistics and Data Science, Yonsei University, Seoul, Republic of Korea. E-mail: ijin@yonsei.ac.kr.

\paragraph{Funding Statement}
This work was partially supported by the National Research Foundation of Korea [grant number NRF-2021S1A3A2A03088949, RS-2023-00217705, RS-2024-00333701; Basic Science Research Program awarded to IHJ].

\paragraph{Competing Interests} None.

\paragraph{Data Availability Statement}
Replication data are available in Harvard Dataverse: \url{https://doi.org/link}.

\paragraph{Ethical Standards}
The research meets all ethical guidelines, including adherence to the legal requirements of the study country.

\paragraph{Author Contributions}
Conceptualization: S.L.; I.K.; J.H.P.; I.H.J. Methodology: I.H.J.; J.H.P.; S.L. Software: S.L. Validation: S.L. Formal analysis: S.L. Data curation: S.L.; I.K. Visualization: S.L. Writing—original draft: S.L.; J.H.P. Writing—review \& editing: S.L.; I.K.; J.H.P.; I.H.J. Supervision: J.H.P.; I.H.J. All authors approved the final submitted draft.

\paragraph{Supplementary Material}
To view supplementary material for this article, please visit \url{https://doi.org/link}.

\printendnotes
\printbibliography

@article{fortunato2010community,
	Author = {Fortunato, Santo},
	Journal = {Physics Reports},
	Number = {3},
	Pages = {75--174},
	Publisher = {Elsevier},
	Title = {Community detection in graphs},
	Volume = {486},
	Year = {2010}}

@article{davis1970expository,
  title={An expository development of a mathematical model of the electoral process},
  author={Davis, Otto A and Hinich, Melvin J and Ordeshook, Peter C},
  journal={American Political Science Review},
  volume={64},
  number={2},
  pages={426--448},
  year={1970},
  publisher={Cambridge University Press}
}

@book{downs1957economic,
  title={An economic theory of democracy},
  author={Downs, Anthony},
  year={1957},
  publisher={Harper \& Row}
}

@article{smith2019geometry,
  title={The geometry of continuous latent space models for network data},
  author={Smith, Anna L and Asta, Dena M and Calder, Catherine A},
  journal={Statistical science: a review journal of the Institute of Mathematical Statistics},
  volume={34},
  number={3},
  pages={428},
  year={2019}
}

@article{nakis2025,
  title={How Low Can You Go? Searching for the Intrinsic Dimensionality of Complex Networks using Metric Node Embeddings},
  author={Nakis, Nikolaos and Holm, Niels Raunkj{\ae}r and Fiehn, Andreas Lyhne and M{\o}rup, Morten},
  journal={arXiv preprint arXiv:2503.01723},
  year={2025}
}

@article{Shin2024,
author = {Sooahn Shin and Johan Lim and Jong Hee Park},
title = {$l$1-based Bayesian Ideal Point Model for Multidimensional Politics},
journal = {Journal of the American Statistical Association},
volume = {120},
number = {550},
pages = {631-644},
year = {2024},
publisher = {ASA Website},
doi = {10.1080/01621459.2024.2425461},
URL = { https://doi.org/10.1080/01621459.2024.2425461},
eprint = {https://doi.org/10.1080/01621459.2024.2425461}}

@article{Clinton2004,
	author = {Clinton, Joshua and Jackman, Simon and Rivers, Douglas},
	journal = {American Political Science Review},
	number = {2},
	pages = {355--370},
	publisher = {Cambridge University Press},
	title = {The Statistical Analysis of Roll Call Data},
	volume = {98},
	year = {2004}}

@article{Barbera2015,
	author = {Barber{\'a}, Pablo},
	journal = {Political Analysis},
	number = {1},
	pages = {76--91},
	publisher = {Cambridge University Press},
	title = {Birds of the same feather tweet together: Bayesian ideal point estimation using Twitter data},
	volume = {23},
	year = {2015}}

@misc{voteview2025,
  author       = {Lewis, Jeffrey B. and Poole, Keith and Rosenthal, Howard and Boche, Adam and Rudkin, Aaron and Sonnet, Luke},
  title        = {Voteview: Congressional Roll-Call Votes Database},
  year         = {2025},
  url          = {https://voteview.com},
  note         = {Data retrieved from Voteview.com}
}

@article{Holland1983,
title = {Stochastic blockmodels: First steps},
journal = {Social Networks},
volume = {5},
number = {2},
pages = {109-137},
year = {1983},
issn = {0378-8733},
doi = {https://doi.org/10.1016/0378-8733(83)90021-7},
url = {https://www.sciencedirect.com/science/article/pii/0378873383900217},
author = {Paul W. Holland and Kathryn Blackmond Laskey and Samuel Leinhardt}
}

@article{jeon:2020,
title = {Mapping Unobserved Item-Respondent Interactions: A Latent Space Item Response Model with Interaction Map},
author = {Jeon, Minjeong and Jin, Ick Hoon and Schweinberger, Michael and Baugh, Samuel},
year = {2021},
journal = {Psychometrika},
volume = {86},
number = 2,
pages = {378-403}
}

@article{gower_generalized_1975,
	title = {Generalized procrustes analysis},
	volume = {40},
	issn = {0033-3123, 1860-0980},
	url = {http://link.springer.com/10.1007/BF02291478},
	doi = {10.1007/BF02291478},
	language = {en},
	number = {1},
	urldate = {2019-10-29},
	journal = {Psychometrika},
	author = {Gower, J. C.},
	year = {1975},
	pages = {33--51}
}

@article{Hoff:2002,
 title = {Latent space approaches to social network analysis},
 journal = {Journal of the American Statistical Association},
 author = {Hoff, P. and Raftery, A. and Handcock, M. S.},
 volume = {97},
 number = 460,
 year = {2002},
 pages = {1090-1098}
}

@article{Handcock:2007,
 title = {Model-based clustering for social network},
 journal = {Journal of the Royal Statistical Society, Series A},
 author = {Handcock, M. S. and Raftery, A. E. and Tantrum, J. M.},
 volume = {170},
 year = {2007},
 pages = {301-354}
}

@article{Krivitsky:2009,
 title = {Representing degree distributions, clustering, and homophily in social networks with latent cluster random network models},
 journal = {Social Networks},
 author = {Krivitsky, P. N. and Handcock, M. S. and Raftery, A. E. and Hoff, P. D.},
 volume = {31},
 year = {2009},
 pages = {204-213}
}

@article{Raftery:2012,
 title = {Fast inference for the latent space network model using a case-control approximate likelihood},
 journal = {Journal of Computational and Graphical Statistics},
 author = {Raftery, A.E. and Niu, X. and Hoff, P.D. and Yeung, K.Y.},
 volume = {21},
 number = 4,
 year = {2012},
 pages = {909-919}
}

@article{Friel:2016,
 title = {Interlocking directorates in {I}rish companies using a latent space model for bipartite networks},
 journal = {Proceedings of the National Academy of Sciences of the United States of America},
 author = {Nial Friel and Riccardo Rastelli and Jason Wyse and Adrian E. Raftery},
 volume = {113},
 year = {2016},
 pages = {6629-6634}
}

@article{sewell2015latent,
  title={Latent space models for dynamic networks},
  author={D. K. Sewell and Y. Chen},
  journal={Journal of the American Statistical Association},
  volume={110},
  pages={1646--1657},
  year={2015},
  publisher={Taylor \& Francis}
}

@article{Schweinberger:03p307,
 author = {Schweinberger, M. and Tom A.B. Snijders},
 title = {Settings in social networks: A measurement model},
 journal = {Sociological Methodology},
 volume = {33},
 year = {2003},
 pages = {307-341}
}

@book{Poole:2007,
  title={Ideology and Congress},
  author={Keith T. Poole and Howard Rosenthal},
  year={2007},
  address={New York},
  publisher={Routledge}
}

@article{Harbridge:2023,
author = {Harbridge-Yong, Laurel and Volden, Craig and Wiseman, Alan},
year = {2023},
pages = {1048-1063},
title = {The Bipartisan Path to Effective Lawmaking},
volume = {85},
journal = {The Journal of Politics},
doi = {10.1086/723805}
}

@article{Clarke:2020,
author = {Clarke, Andrew J.},
title = {Party Sub-Brands and American Party Factions},
journal = {American Journal of Political Science},
volume = {64},
number = {3},
pages = {452-470},
doi = {https://doi.org/10.1111/ajps.12504},
url = {https://onlinelibrary.wiley.com/doi/abs/10.1111/ajps.12504},
eprint = {https://onlinelibrary.wiley.com/doi/pdf/10.1111/ajps.12504},
year = {2020}
}

@article{Duck-Mayr:2023, title={Ends Against the Middle: Measuring Latent Traits when Opposites Respond the Same Way for Antithetical Reasons}, volume={31}, DOI={10.1017/pan.2022.33}, number={4}, journal={Political Analysis}, author={Duck-Mayr, JBrandon and Montgomery, Jacob}, year={2023}, pages={606–625}}

@article{Hoff:2007,
  title={Modeling homophily and stochastic equivalence in symmetric relational data},
  author={Hoff, Peter},
  journal={Advances in neural information processing systems},
  volume={20},
  year={2007}
}

@article{mccarty2016,
	title = {In Defense of {DW}-{NOMINATE}},
	volume = {30},
	copyright = {https://www.cambridge.org/core/terms},
	issn = {0898-588X, 1469-8692},
	url = {https://www.cambridge.org/core/product/identifier/S0898588X16000110/type/journal_article},
	doi = {10.1017/S0898588X16000110},
	language = {en},
	number = {2},
	urldate = {2025-08-31},
	journal = {Stud. Am. Pol. Dev.},
	author = {McCarty, Nolan},
	year = {2016},
	pages = {172--184},
}

@article{Hoff:2009,
	title = {Multiplicative latent factor models for description and prediction of social networks},
	volume = {15},
	issn = {1572-9346},
	url = {https://doi.org/10.1007/s10588-008-9040-4},
	doi = {10.1007/s10588-008-9040-4},
	language = {en},
	number = {4},
	urldate = {2025-09-01},
	journal = {Comput Math Organ Theory},
	author = {Peter Hoff},
	year = {2009},
	pages = {261--272},
}

@inproceedings{Li:2011,
	address = {Barcelona, Catalonia, Spain},
	series = {{IJCAI}'11},
	title = {Generalized latent factor models for social network analysis},
	isbn = {978-1-57735-514-4},
	urldate = {2025-08-31},
	booktitle = {Proceedings of the {Twenty}-{Second} international joint conference on {Artificial} {Intelligence} - {Volume} {Volume} {Two}},
	publisher = {AAAI Press},
	author = {Li, Wu-Jun and Yeung, Dit-Yan and Zhang, Zhihua},
	year = {2011},
	pages = {1705--1710},
}

@article{Airoldi:2008,
	title = {Mixed Membership Stochastic Blockmodels},
	volume = {9},
	issn = {1533-7928},
	url = {http://jmlr.org/papers/v9/airoldi08a.html},
	number = {65},
	urldate = {2025-09-01},
	journal = {Journal of Machine Learning Research},
	author = {Airoldi, Edoardo M. and Blei, David M. and Fienberg, Stephen E. and Xing, Eric P.},
	year = {2008},
	pages = {1981--2014}
}

@article{Nowicki:2001,
	title = {Estimation and Prediction for Stochastic Blockstructures},
	volume = {96},
	issn = {0162-1459},
	url = {https://www.jstor.org/stable/2670253},
	number = {455},
	urldate = {2025-09-01},
	journal = {Journal of the American Statistical Association},
	author = {Nowicki, Krzysztof and Snijders, Tom A. B.},
	year = {2001},
	pages = {1077--1087},
}

@inproceedings{young:2007,
	address = {Berlin, Heidelberg},
	title = {Random Dot Product Graph Models for Social Networks},
	isbn = {978-3-540-77004-6},
	doi = {10.1007/978-3-540-77004-6\_11},
	language = {en},
	booktitle = {Algorithms and {Models} for the {Web}-{Graph}},
	publisher = {Springer},
	author = {Young, Stephen J. and Scheinerman, Edward R.},
	editor = {Bonato, Anthony and Chung, Fan R. K.},
	year = {2007},
	pages = {138--149},
}

@inproceedings{Kemp:2006,
	address = {Boston, Massachusetts},
	series = {{AAAI}'06},
	title = {Learning systems of concepts with an infinite relational model},
	isbn = {978-1-57735-281-5},
	urldate = {2025-08-31},
	booktitle = {Proceedings of the 21st national conference on {Artificial} intelligence - {Volume} 1},
	publisher = {AAAI Press},
	author = {Kemp, Charles and Tenenbaum, Joshua B. and Griffiths, Thomas L. and Yamada, Takeshi and Ueda, Naonori},
	year = {2006},
	pages = {381--388},
}

@Manual{pscl2024,
    title = {{pscl}: Classes and Methods for {R} Developed in the Political Science Computational Laboratory},
    author = {Simon Jackman},
    organization = {University of Sydney},
    address = {Sydney, Australia},
    year = {2024},
    note = {R package version 1.5.9},
    url = {https://github.com/atahk/pscl/},
  }

@Article{latentnet2008,
    title = {Fitting position latent cluster models for social networks with latentnet},
    author = {Pavel N. Krivitsky and Mark S. Handcock},
    year = {2008},
    volume = {24},
    number = {5},
    journal = {Journal of Statistical Software},
    doi = {10.18637/jss.v024.i05},
  }

@article{Hoff:2021,
 ISSN = {08834237, 21688745},
 URL = {https://www.jstor.org/stable/26997946},
 author = {Peter Hoff},
 journal = {Statistical Science},
 number = {1},
 pages = {pp. 34--50},
 publisher = {Institute of Mathematical Statistics},
 title = {Additive and Multiplicative Effects Network Models},
 urldate = {2025-09-12},
 volume = {36},
 year = {2021}
}

@inproceedings{Palla:2012,
author = {Palla, Konstantina and Knowles, David A. and Ghahramani, Zoubin},
title = {An infinite latent attribute model for network data},
year = {2012},
isbn = {9781450312851},
publisher = {Omnipress},
address = {Madison, WI, USA},
booktitle = {Proceedings of the 29th International Coference on International Conference on Machine Learning},
pages = {395–402},
numpages = {8},
location = {Edinburgh, Scotland},
series = {ICML'12}
}

@INPROCEEDINGS{Morup:2011,
  author={Mørup, Morten and Schmidt, Mikkel N. and Lars Kai Hansen},
  booktitle={2011 IEEE International Workshop on Machine Learning for Signal Processing},
  title={Infinite multiple membership relational modeling for complex networks},
  year={2011},
  volume={},
  number={},
  pages={1-6},
  doi={10.1109/MLSP.2011.6064546}}

@inproceedings{zhou2015infinite,
  title={Infinite edge partition models for overlapping community detection and link prediction},
  author={Zhou, Mingyuan},
  booktitle={Artificial intelligence and statistics},
  pages={1135--1143},
  year={2015},
  organization={PMLR}
}

@inproceedings{miller:2009,
title = {Nonparametric Latent Feature Models for Link Prediction},
volume = {22},
url = {https://proceedings.neurips.cc/paper_files/paper/2009/file/437d7d1d97917cd627a34a6a0fb41136-Paper.pdf},
booktitle = {Advances in {Neural} {Information} {Processing} {Systems}},
publisher = {Curran Associates, Inc.},
author = {Miller, Kurt and Jordan, Michael and Griffiths, Thomas},
editor = {Bengio, Y. and Schuurmans, D. and Lafferty, J. and Williams, C. and Culotta, A.},
year = {2009},
}

@article{daudin:2008,
title = {A mixture model for random graphs},
volume = {18},
issn = {1573-1375},
url = {https://doi.org/10.1007/s11222-007-9046-7},
doi = {10.1007/s11222-007-9046-7},
number = {2},
journal = {Statistics and Computing},
author = {Daudin, J.-J. and Picard, F. and Robin, S.},
year = {2008},
pages = {173--183},
}

@article{Vu:2013,
title = {Model-based clustering of large networks},
volume = {7},
url = {https://doi.org/10.1214/12-AOAS617},
doi = {10.1214/12-AOAS617},
number = {2},
journal = {The Annals of Applied Statistics},
author = {Vu, Duy Q. and Hunter, David R. and Schweinberger, Michael},
year = {2013},
pages = {1010-1039},
}

@article{Karrer:2011,
title = {Stochastic blockmodels and community structure in networks},
author = {Karrer, Brian and Newman, M. E. J.},
journal = {Phys. Rev. E},
volume = {83},
issue = {1},
pages = {016107},
numpages = {10},
year = {2011},
publisher = {American Physical Society},
doi = {10.1103/PhysRevE.83.016107},
url = {https://link.aps.org/doi/10.1103/PhysRevE.83.016107}
}

@article{binding2023non,
  title={Non-separable preferences in the statistical analysis of roll call votes},
  author={Binding, Garret and Stoetzer, Lukas F},
  journal={Political Analysis},
  volume={31},
  number={3},
  pages={352--365},
  year={2023},
  publisher={Cambridge University Press}
}

@article{kim2018estimating,
  title={Estimating spatial preferences from votes and text},
  author={Kim, In Song and Londregan, John and Ratkovic, Marc},
  journal={Political Analysis},
  volume={26},
  number={2},
  pages={210--229},
  year={2018},
  publisher={Cambridge University Press}
}

@article{moser2021multiple,
  title={Multiple ideal points: Revealed preferences in different domains},
  author={Moser, Scott and Rodr{\'i}guez, Abel and Lofland, Chelsea L},
  journal={Political Analysis},
  volume={29},
  number={2},
  pages={139--166},
  year={2021},
  publisher={Cambridge University Press}
}

@article{lei2025novel,
  title={A novel class of unfolding models for binary preference data},
  author={Lei, Rayleigh and Rodriguez, Abel},
  journal={Political Analysis},
  volume={33},
  number={1},
  pages={32--48},
  year={2025},
  publisher={Cambridge University Press}
}

@article{lipman2025explaining,
  title={Explaining differences in voting patterns across voting domains using hierarchical bayesian models},
  author={Lipman, Erin and Moser, Scott and Rodriguez, Abel},
  journal={Political Analysis},
  pages={1--21},
  year={2025},
  publisher={Cambridge University Press}
}

@article{marble2022structure,
  title={The structure of political choices: Distinguishing between constraint and multidimensionality},
  author={Marble, William and Tyler, Matthew},
  journal={Political Analysis},
  volume={30},
  number={3},
  pages={328--345},
  year={2022},
  publisher={Cambridge University Press}
}

@article{lo2025statistical,
  title={A Statistical Model of Bipartite Networks: Application to Cosponsorship in the United States Senate},
  author={Lo, Adeline and Olivella, Santiago and Imai, Kosuke},
  journal={Political Analysis},
  pages={1--20},
  year={2025},
  publisher={Cambridge University Press}
}

@article{Diaconis2008,
author = {Diaconis, Persi and Goel, Sharad and Holmes, Susan},
year = {2008},
pages = {},
title = {Horsehoes in multidimensional scaling and local kernel methods},
volume = {2},
journal = {The Annals of Applied Statistics},
doi = {10.1214/08-AOAS165}
}
\end{document}